\documentstyle[12pt,preprint,aps,epsf,floats]{revtex}
\def\NPB#1#2#3{Nucl. Phys. {\bf B#1}, #3 (19#2)}

\def\PLB#1#2#3{Phys. Lett. {\bf B#1}, #3 (19#2)}
\def\PLBMM#1#2#3{Phys. Lett. {\bf B#1}, #3 (20#2)}
\def\PLBold#1#2#3{Phys. Lett. {\bf#1B}, #3 (19#2)}
\def\PRD#1#2#3{Phys. Rev. {\bf D#1}, #3 (19#2)}
\def\PRDMM#1#2#3{Phys. Rev. {\bf D#1}, #3 (20#2)}
\def\PRL#1#2#3{Phys. Rev. Lett. {\bf#1}, #3 (19#2)}
\def\PRLMM#1#2#3{Phys. Rev. Lett. {\bf#1}, #3 (20#2)}
\def\PRep#1#2#3{Phys. Rep. {\bf#1}, #3 (19#2)}

\newcommand{\PSbox}[3]{\mbox{\rule{0in}{#3}
\includegraphics{#1}\hspace{#2}}}

\newcommand{\mweak}{M_{\text{Weak}}}
\newcommand{\mplanck}{M_{\text{Planck}}}
\newcommand{\hc}{\text{H.c.}}

\newcommand{\mrU}{{U}^{\prime}}
\newcommand{\mrD}{{D}^{\prime}}
\newcommand{\mrQ}{{Q}^{\prime}}
\newcommand{\mrL}{{L}^{\prime}}
\newcommand{\mrE}{{E}^{\prime}}

\newcommand{\mrQthird}{{{Q}^{\prime}}_3}
\newcommand{\sU}{{\widetilde{U}}}
\newcommand{\sD}{{\widetilde{D}}}
\newcommand{\sQ}{{\widetilde{Q}}}
\newcommand{\sL}{{\widetilde{L}}}
\newcommand{\sE}{{\widetilde{E}}}

\newcommand{\mrsU}{{\widetilde{U^{\prime}}}}
\newcommand{\mrsD}{{\widetilde{D^{\prime}}}}
\newcommand{\mrsQ}{{\widetilde{Q^{\prime}}}}
\newcommand{\mrsL}{{\widetilde{L^{\prime}}}}
\newcommand{\mrsE}{{\widetilde{E^{\prime}}}}

\newcommand{\mru}{{u}^{\prime}}
\newcommand{\mruL}{{u}^{\prime}_{L}}
\newcommand{\mruR}{{u}^{\prime}_{R}}
\newcommand{\mrd}{{d}^{\prime}}

\newcommand{\mrq}{{q}^{\prime}}
\newcommand{\mrt}{{t}^{\prime}}
\newcommand{\mrtL}{{t}^{\prime}_{L}}
\newcommand{\mrtR}{{t}^{\prime}_{R}}
\newcommand{\mrb}{{b}^{\prime}}

\newcommand{\mrl}{{l}^{\prime}}
\newcommand{\mre}{{e}^{\prime}}

\newcommand{\hu}{{H}_{2}}
\newcommand{\hd}{{H}_{1}}
\newcommand{\mrhu}{{{H}^{\prime}}_{2}}
\newcommand{\mrhd}{{{H}^{\prime}}_{1}}
\newcommand{\hinou}{{\widetilde{H}}_{2}}
\newcommand{\hinod}{{\widetilde{H}}_{1}}
\newcommand{\mrhinou}{{\widetilde{H}}^{\prime}_{2}}
\newcommand{\mrhinod}{{\widetilde{H}}^{\prime}_{1}}
\newcommand{\bino}{\widetilde{B}}
\newcommand{\wino}{\widetilde{W}}
\newcommand{\gluino}{\widetilde{g}}
\newcommand{\adjoint}{\Phi_{V}}
\newcommand{\sadjoint}{\phi_{V}}
\newcommand{\fadjoint}{\psi_{V}}

\newcommand{\gadjoint}{\Phi_{g}}
\newcommand{\gsadjoint}{\phi_{g}}
\newcommand{\gfadjoint}{\psi_{g}}

\newcommand{\wadjoint}{\Phi_{W}}
\newcommand{\wsadjoint}{\phi_{W}}
\newcommand{\wfadjoint}{\psi_{W}}

\newcommand{\badjoint}{\Phi_{B}}
\newcommand{\bsadjoint}{\phi_{B}}
\newcommand{\bfadjoint}{\psi_{B}}

\tighten
\begin{document}
\preprint{
\noindent
\begin{minipage}[t]{3in}
\begin{flushleft}
June 2000 \\
\end{flushleft}
\end{minipage}
\hfill
\begin{minipage}[t]{3in}
\begin{flushright}
MIT--CTP--2987\\
hep-ph/0006174\\
\vspace*{.7in}
\end{flushright}
\end{minipage}
}

\draft

\title{
Low-Energy Limits of Theories With Two Supersymmetries 
}

\author{Nir Polonsky and Shufang Su 
\vspace*{.2in}
}
\address{
Center for Theoretical Physics, Massachusetts Institute
of Technology\\
Cambridge, MA 02139 USA
\vspace*{.2in}}


\maketitle

\begin{abstract}
Given its non-renormalization properties,
low-energy supersymmetry provides an attractive 
framework for extending the Standard Model
and for resolving the hierarchy problem. 
Models with softly broken $N =1$ supersymmetry 
were extensively studied and are phenomenologically successful. 
However, it could be that an extended $N = 2$ supersymmetry 
survives to low energies, as suggested by various constructions.
We examine the  phenomenological viability and 
implications of such a scenario.
We show that consistent chiral fermion mass generation 
emerges in $N=2$ theories, which are vectorial, as 
a result of supersymmetry breaking at low energies.
A rich mirror quark and lepton spectrum near the 
weak scale with model-dependent decay modes is predicted.
A $Z_{2}$ mirror parity is shown to 
play an important role in determining the phenomenology of the models.
It leads, if conserved, to a new stable particle, the LMP.
Consistency  of the $N=2$ framework and its unique spectrum with 
electroweak precision data is considered, and the 
discovery potential in the next generation of 
hadron collider experiments is stressed.
Mirror quark pair production provides the most promising discovery channel. 
Higgs searches are also discussed and 
it is shown that there is no upper bound
on the  prediction for the Higgs boson mass in 
the framework of low-energy supersymmetry breaking, in general, and
in the $N=2$ framework, in particular.
Possible $N = 2$ realizations of flavor symmetries and of neutrino masses
are also discussed.
\end{abstract}

\pacs{PACS numbers: 12.60.Jv, 11.30.Pb, 14.80.-j}


%
\section{Introduction}
\label{sec:introduction}
%

Supersymmetry and its boson-fermion
symmetry provide an attractive framework 
for embedding  the standard model
of electroweak and strong interactions (SM) \cite{susy}.
The electroweak scale is understood in this framework
as roughly the scale of supersymmetry breaking in 
the global theory and is protected, in general, 
from destabilization at the quantum level.
In particular, softly broken $N = 1$ supersymmetry provides a 
phenomenologically successful extension of the SM \cite{mssm}.
The particle content is the minimal one required 
by the boson-fermion symmetry and, regardless
of the exact details of the soft supersymmetry breaking
(SSB) spectrum parameters, the corresponding $\beta$-functions predict
gauge coupling unification at a scale of 
${\cal{O}}(10^{16})$ GeV \cite{unification}.
The theory tends to decouple from most electroweak observables
(for sparticles of ${\cal{O}}(300)$ GeV) \cite{decouple}
while the absence of flavor changing neutral currents 
is a source of information about the  high-energy origins of 
the low-energy effective theory.
At high energies, the rigid supersymmetry can be extended to
supergravity \cite{susy,mssm}: 
The first step towards gravity-gauge unification and further
embedding of the SM in a theory of quantum gravity of which supergravity
is the sub-Planckian limit, for example, superstring theory.
%

However, there exists a tension between a ``bottom-up''
approach, which beginning with the SM motivates 
its $N = 1$ supersymmetry extension, and a ``top-down'' approach, 
which beginning
with a superstring theory often suggests that an extended $N = 2$ supersymmetry
is broken at some energy directly to $N = 0$ \cite{breaking}. 
If supersymmetry is to stabilize the weak scale 
and resolve the hierarchy problem associated with its instability
in the SM, then the extended
supersymmetry can be broken in this case only near that scale.
Indeed, current knowledge of string theory is far from sufficient for
understanding its electroweak-scale limit or how 
the SM would be embedded in such a theory, and therefore 
it is premature to draw conclusions from string 
theory regarding the nature of the weak-scale supersymmetry.  
Nevertheless, it is an intriguing and highly 
interesting question to ask whether an $N = 2$ 
supersymmetry extension of the SM at weak-scale energies
is phenomenologically viable, 
what constraints the infra-red SM limit imposes on ultra-violet realizations
of such a theory, and what would be its signatures.
Here, as a first step towards addressing these questions, 
we will investigate some of their more fundamental
aspects, laying the foundation for, 
and hopefully intriguing, further discussion. 
%

The phenomenology of $N=2$ supersymmetry and its extended spectrum were
studied over the years by only a couple of groups \cite{west,GPZ,fayet}. 
Its ``ultra-violet'' elegance stemming from its constrained structure 
(for example, there is  only one coupling in the theory, the gauge coupling,
and the theory is not renormalized beyond one loop)
does not translate to an equivalent elegance in the infra-red. 
On the contrary, the $N=2$ intrinsic constraints
make it difficult to reconcile the framework
with the SM and with observations.
Most notably, the theory does not contain chiral 
Yukawa couplings or any other source of chiral mass generation. 
Once supersymmetry is broken the fermion mass issue can be resolved.
However, one then finds that the $N=2$ theory does not decouple from
electroweak observables (nor does it suggest gauge unification).
These issues place strong constraints on the properties of the
extended spectrum that $N=2$ supersymmetry predicts.
%

Theories with two supersymmetries contain a rich spectrum:
While each SM fermion (boson) is accompanied by a boson (fermion)
superpartner to form a chiral or a vector superfield in the $N = 1$ extension,
each chiral $N = 1$ superfield is further accompanied by an anti-chiral 
superfield to form a vector-like hypermultiplet in the $N = 2$ extension.
An $N = 1$ vector superfield is accompanied in the $N = 2$ 
extension by an $N=1$ chiral superfield  in the appropriate representation,
the {\it mirror} gauge superfield.
For example, a SM quark is partnered, in addition to the squark, also
with a {\it mirror} quark and a {\it mirror} squark. The gauge boson
is partnered not only with the gaugino but also with a complex scalar
and additional Majorana fermion in the adjoint representation 
(or singlets in the Abelian case).
The number of particles is increased four times with respect to the SM!
%

In describing the extended spectrum we used the $N=1$ superfield language.
Indeed, it is possible (and we will do so) to formulate
the $N=2$ framework in this language. In order to impose the additional
$N=2$ constraints one has to specify a set of global $R$ symmetries.
It includes a vectorial $SU(2)_{R}$ exchange $R$-symmetry which forbids, 
as mentioned above, any chiral fermion Yukawa or mass terms.
Once the vectorial symmetry is broken,
all chiral and anti-chiral fermion masses are proportional to the
Higgs vacuum expectation values (VEVs) which spontaneously break the SM
$SU(2)_{L}$.  One expects that the 
new particles are entangled with the ordinary SM fields,
as the gauge symmetries do not forbid their mixing.
This again provides an important set of constraints on the theory
and on the dynamics that breaks it.
It also provides clear tests of the framework.
Most importantly, unlike a $N =1$ theory that must be discovered
via its somewhat arbitrary predictions for the  spectrum of
new bosons and Majorana fermions,   
an $N = 2$ theory would be readily discovered or
excluded in the next generation of hadron collider via its strongly
constrained predictions of the mirror (anti-chiral) fermion spectrum,
which is not expected to be much heavier than the top quark.
Henceforth, a study of the $N=2$ framework is
timely and well motivated.
%

The knowledgeable reader may be questioning the validity
of any such an extension which  contains 
contributions of three anti-chiral families to the oblique
$S$ parameter \cite{S} (which measures quantum corrections 
from new physics to $Z - \gamma$ mixing).
Usually, one assumes that mass-dependent terms are negligible,
as is appropriate in the decoupling limit $m_{f_{\rm new}} \gg m_{Z}$.
In that case, the mass-independent contribution
of each chiral or anti-chiral generation
to $S$ is positive, contrary to
current measurements which imply $S \leq  0$ \cite{pdg}.
Therefore, one may argue that $S$ excludes extra (anti-)chiral families.
However, the $N=2$ spectrum is far too rich and complicated 
to allow for such arguments. 
Current data does not exclude (though it does not suggest)
three anti-chiral families
if the spectrum of the $N=2$ mirror fermions 
breaks the custodial $SU(2)$ symmetry of the electroweak interactions 
and is (at least partially) ``light'' \cite{oblique1}.
The situation is even more arbitrary if the 
Majorana fermion spectrum, which also contains
custodial $SU(2)$ breaking mass terms, is considered \cite{oblique2}.
We therefore proceed and investigate $N = 2$ models, further discussing
this and other phenomenological issues in a dedicated section. 
Our main focus, however, is establishing tools 
for the construction of the chiral fermion spectrum
below the $N = 2$ breaking scale.
%

The fermion mass problems in these models
has many facets. First and foremost, the generation of any chiral
spectrum must be a result of supersymmetry breaking.
Secondly, the two sectors have to be distinguished
with sufficiently heavy mirror fermions and (relatively)
light ordinary fermions, with any mixing between the two sectors
suppressed, at least in the case of the first two families.
In addition there are the issues of the heavy third family
and of the very light neutrinos in the ordinary sector,
and subsequently, of flavor symmetries and their relation
to  supersymmtry breaking.  
In order to address these issues 
we choose to formulate a global $N =2$ theory
as an $N = 1$ theory with a second supersymmetry
manifest only through global $R$-symmetries which are imposed
on the $N = 1$ description. 
(For example, the $SU(2)_{R}$ mentioned above.)
This is a standard procedure \cite{west}
that allows, in our case,  the usage of the $N = 1$ 
spurion formalism \cite{GG} in the construction of the fermion spectrum. 
Specifically, we assume below that
\begin{itemize}
\item 
The matter content is that of the minimally extended
$N=2$ supersymmetric SM (MN2SSM) 
(and that of a flavor sector, if exists) given in terms
of $N=1$ chiral and vector superfields.
\item 
The $N=2$ imposed global symmetries (or a subset thereof)
are explicitly broken by non-renormalizable terms in the Kahler
potential. These terms are characterized by a scale $M$. 
In the limit $M \rightarrow \infty$
the full supersymmetry is restored.
\item
The only VEVs are  ($N= 1$ breaking) $F$-type VEVs 
which generate all dimensionful and dimensionless 
couplings in the electroweak theory (aside from the gauge coupling).
Electroweak symmetry breaking VEVs
(and flavor symmetry breaking VEVs) are then induced in
the resulting effective theory.
\item
To the most part we will also assume that
some flavor and ``mirror'' symmetries, which do not commute
with the $N=2$ $R$-symmetries, are conserved in the resulting effective
theory.
\end{itemize}
We will explore the chiral fermion spectrum within this framework
and establish phenomenologically
viable  low-energy limits of the $N=2$ framework
which could be probed, given the fermion spectrum, in the near future. 
%

We briefly review $N = 2$ supersymmetry
in Section~\ref{sec:theory}. Though we do not focus on the boson spectrum
and the dimensionful SSB parameters,
they are straightforward to write in the $N=1$ formalism. This will
be done as a warm-up exercise
in Section~\ref{sec:soft} (and again, using the spurion formalism,
in Section~\ref{sec:tree}). The softly broken $N = 2$ model
will be compared to its $N =1$ equivalent.
In Section~\ref{sec:loop} we consider
the possibility of breaking the chiral symmetries
primarily in the SSB scalar potential, 
generating the fermion spectrum only radiatively \cite{west,BFPT}.
While this option is viable in some cases for the SM spectrum, 
it cannot provide a consistent mirror fermion spectrum. 
In Section~\ref{sec:hard} we exploit
the $N = 1$ spurion formalism to classify the most general supersymmetry
breaking framework, and in Section~\ref{sec:tree} we use it
to discuss a general $N=2$ framework and 
the possible tree-level origins of the
chiral fermion spectrum. We find that certain Kahler operators
can generate such a spectrum as long as the supersymmtries are broken at 
relatively low-energies, which we will assume.
(Note that large parts of our discussion are applicable to 
low-energy $N = 1$ supersymmetry breaking as well.) 
The special issue of the heavy SM third family fermions 
is addressed in Section~\ref{sec:mixing}.
In Section~\ref{sec:neutrino} we comment on neutrino physics within
the context of $N=2$ supersymmetry. Direct and indirect signatures
and other phenomenological issues
are discussed in Section~\ref{sec:pheno}. 
Unlike $N=1$ supersymmetry, $N =2$ cannot escape
detection in the next generation of hadron colliders!
We conclude with a summary, an outlook,
and a comparison to previous constructions, in Section~\ref{sec:conclusions}.
%
%
%


%
\section{The MN2SSM Framework}
\label{sec:theory}
%

The  N=2 supersymmetry algebra has two spinorial generators 
$Q^i_{\alpha}$, $i=1,2$, 
satisfying 
\begin{equation}
\{{Q}_{\alpha}^{i}, {{\bar{Q}}}_{\dot{\alpha}}^{j}\}=
\sigma^{\mu}_{\alpha\dot{\alpha}}P_{\mu}\delta^{ij},
\end{equation}
where $\sigma^{\mu}$ are, as usual, the Pauli matrices 
and $P_{\mu}$ is the momentum. The supercharges ${Q}_{\alpha}^{i}$
form a doublet of the (exchange) 
$SU(2)_{R}$ $R$-symmetry, which must be imposed
when using the $N=1$ formulation to describe a $N=2$ theory.
The lowest $N =2$ spin representations, which are the relevant 
ones for embedding the SM, 
are the hypermultiplet and vector multiplet. 
Written in the familiar $N=1$ language, the  hypermultiplet is 
composed of two $N=1$ chiral multiplets $X=(x,\ \psi_x)$ and $Y=(y,\ \psi_y)$,
with $Y$ occupying  representations ${\cal{R}}$ of the gauge groups
which are conjugate to that of $X$, ${\cal{R}}(X) = {\cal{R}}(Y^{\dagger})$.
Schematically, the hypermultiplet is described by a ``diamond'' plot
\begin{center}
\begin{tabular}{cccccccc}
&&$\psi_{x}$ &&&&& $+\frac{1}{2}$\\
&$\nearrow$&&&&&&\\
$x$ &&&& $y^{\dagger}$&&&$0$ \\
&&&$\swarrow$&&&&\\
&&$\bar{\psi}_{y}$&&&&&$-\frac{1}{2}$ 
\end{tabular}
\end{center}
where the first, second and third rows
correspond to helicity $-1/2$, $0$, and $+1/2$ states.
The vector multiplet contains a  $N=1$ vector multiplet 
$V=(V^{\mu}, \lambda)$, where  $\lambda$ is a gaugino,
and a $N=1$ chiral multiplet $\adjoint=(\sadjoint,\ 
\fadjoint)$ in the adjoint representation of the gauge group (or a singlet
in the Abelian case).  Schematically, it is described by
\begin{center}
\begin{tabular}{cccccccc}
&&$V^{\mu}$ &&&&& $1$ \\
&$\nearrow$&&&&&&\\
$\lambda$ &&&& $\fadjoint$&&&$\frac{1}{2}$ \\
&&&$\nearrow$&&&\\
&&$\sadjoint$ &&&&& $0$
\end{tabular}
\end{center}
where the first, second and third rows
correspond to helicity $0$, $1/2$, and $1$ states.
The $N=1$ superfields are given by the two $45^{\circ}$ 
sides of each diamond (indicated by arrows),
with the gauge field arranging itself in its chiral representation
$W_{\alpha} \sim \lambda_{\alpha} +\theta_{\alpha}V$.
The particle content doubles in comparison to the $N=1$ supersymmetry case
and it is four times that of the SM. 
For each of the usual chiral fermions $\psi_{x}$ and
its complex-scalar partner $x$, there are a conjugate
mirror fermion $\psi_{y}$  and complex scalar  $y$
(so that the theory is vectorial).
For each gauge boson and 
gaugino, there is a {\it mirror} gauge boson  $\sadjoint$ and 
a {\it mirror} gaugino $\fadjoint$.  
%

The $N=0$ boson and fermion components of the 
hyper and vector-multiplet form $SU(2)_{R}$ representations.
States with equal helicity form a $SU(2)_{R}$ doublet $(x,y^{\dagger})$
and an anti-doublet $(\fadjoint,\lambda)$, 
while all other states are $SU(2)_{R}$ singlets.
In fact, the full $R$-symmetry is $U(2)_{R}$
of which the exchange $SU(2)_{R}$ is a subgroup. 
There are additional $U(1)_{R}^{N=2}$, 
$U(1)_{J}^{N=2}$ subgroups such that the $R$-symmetry is either
$SU(2)_{R}\times U(1)_{R}^{N=2}$, or in some cases only
a reduced $U(1)_{J}^{N=2}\times U(1)_{R}^{N=2}$.
The different superfields $X \sim x +\theta\psi_{x}$, etc.
transform under the $U(1)$ 
symmetries with charges $R$ and $J$ given by
\begin{equation}
R({X}) = r = -R({Y}), \ \ \ R({\adjoint}) = -2, \ \ \ 
\end{equation}
\begin{equation}
J({X}) = -1 = J({Y}), \ \ \ J({\adjoint}) = 0, \ \ \ 
\end{equation}
and $R(W_{\alpha}) = J(W_{\alpha}) = -1$.
The (manifest) supercoordinate $\theta$ (which carries a chiral index,
denoted explicitly in some cases)
has, as usual, charge $R({\theta}) = J({\theta}) = -1$. 
%

The $SU(2)_{R}\times U(1)_{R}^{N=2}$ invariant $N=2$ Lagrangian can be written 
in the $N=1$ language as \cite{susy,west}
\begin{eqnarray}
L &=&\frac{1}{8g^2}[W^{\alpha}W_{\alpha}]_F+[\sqrt{2}igY\adjoint{X}]_F+\hc
\nonumber \\
&&
+[2{\text {Tr}}(
\adjoint^{\dagger}{\text{e}}^{2gV}\adjoint{\text{e}}^{-2gV}+X^{\dagger}
{\text{e}}^{2gV}X+Y^{\dagger}{\text{e}}^{-2gV^T}Y)]_D,
\label{L}
\end{eqnarray}
where $\adjoint=\Phi^a_VT^a$ and $V=V^aT^a$, 
$T^a$ being the respective generators.
The second $F$-term is the superpotential. 
The only free coupling is the gauge coupling constant $g$: The coupling 
constant of the Yukawa term in the superpotential is 
fixed by the gauge coupling due to a global $SU(2)_{R}$. 
In particular, the $SU(2)_{R}$ symmetry forbids any chiral Yukawa terms
so that fermion mass generation is linked to supersymmetry breaking, as 
will be discussed in the following sections.
Note that the $U(1)_{R}^{N=2}$ forbids any mass terms $W\sim \mu^{\prime} XY$
(and the full $R$-symmetry forbids the usual $N=1$ $\mu$-term 
$W\sim \mu \hd\hu$ to be discussed later).
Unlike the $SU(2)_{R}$, $U(1)_{R}^{N=2}$ can survive supersymmetry breaking.
%

The $N=2$ Lagrangian (\ref{L}) contains
several discrete symmetries, 
which  may or may not be broken in the broken supersymmetry regime.
There is a trivial 
extension of the usual $N=1$ $R$-parity ($R_{P}$) 
$Z_{2}$ symmetry which does not distinguish the 
ordinary fields from their mirror partners: 
\begin{equation}
\theta\rightarrow{-}\theta, \ \ \ X_M\rightarrow{-}X_M, \ \ \ 
Y_M\rightarrow{-}Y_M,
\end{equation}
where all other supermultiplets are $R_{P}$-even and
where the hypermultiplets have been divided into the odd matter multiplets
$(X_M,\ Y_M)$ and the even Higgs multiplets $(X_H,\ Y_H)$.
(Note that $V$ is even but $W_{\alpha}$ is odd.)
As in the $N=1$ case, all the ordinary and mirror quarks, leptons and Higgs 
bosons are $R_{P}$-even, while the ordinary and mirror
gauginos are $R_{P}$-odd. $R_{P}$ is conveniently used to define
the superpartners (or sparticles) as the $R_{P}$-odd particles \cite{FF}.  
The lightest  superparticle (LSP) is stable if $R_{P}$ remains unbroken.
%

A second parity, called mirror parity ($M_{P}$), 
distinguishes the mirror particles
from their partners:
\begin{equation}\theta\rightarrow \theta, \ \ \
Y_M\rightarrow{-}Y_M, \ \ \ Y_H\rightarrow{-}Y_H, \ \ \ 
\adjoint\rightarrow{-}\adjoint,
\end{equation}
and all other superfields (including $W_{\alpha}$) are $M_{P}$-even.
It is convenient to use mirror parity to define the mirror particles
as the $M_{P}$-odd particles. (This definition should not be confused
with other definitions of mirror particles 
used in the literature and which are
based on a left-right group
$SU(2)_{L} \times SU(2)_{R}$ or a mirror world which interacts
only gravitationally with the SM world.)
The lightest mirror parity odd particle (LMP) 
is also stable in a theory with 
unbroken mirror parity.  However, if supersymmetry breaking
does not preserve mirror parity,  
mixing between the ordinary matter and the mirror fields is allowed. 
%

There is also a reflection (exchange) symmetry 
(which must be broken at low energies), the 
mirror exchange symmetry:
\begin{equation}
X\leftrightarrow{Y},\ \ \ \adjoint\leftrightarrow\adjoint^{\text{T}}, \ \ \ 
V\leftrightarrow{-}V^{\text{T}}.
\end{equation}
Like in the case of this continuous $SU(2)_{R}$,
if the reflection symmetry remains exact after supersymmetry breaking
then for each left-handed fermion  there would be a degenerate 
right handed mirror fermion in the conjugate gauge 
representation, which is phenomenologically not acceptable.
%

For easy reference, we list 
in Table~\ref{table:content} the minimal particle content of 
the MN2SSM, where a mirror partner $Y$ ($\Phi_{V}$)
exists for every ordinary superfield $X$ ($V$) of the  Minimal 
$N = 1$ Supersymmetric extension of the Standard Model (MSSM). 
We could eliminate one Higgs hypermultiplet and treat $\hd$ and $\hu$ as 
mirror partners.  However, this could lead to the spontaneous breaking of  
mirror parity when the Higgs bosons acquire VEVs, 
and as a result, to a more complicated radiative structure than 
the theory with two Higgs hypermultiplets. 
(We note, however, that it is possible in these theories
that only one Higgs doublet acquires a VEV as the chiral Yukawa coupling
are related to supersymmetry breaking and, unlike in the $N = 1$ case
with high-energy supersymmetry breaking, 
are not necessarily constrained by holomorphicity.)
%

\begin{table}
\caption{Hypermultiplets and vector multiplets in the MN2SSM.}
\label{table:content}
\begin{tabular}{c|c|c|c|c|c|c}
&\multicolumn{3}{c|}{$X,\,\,V$}&\multicolumn{3}{c}{$Y,\,\,\Phi_{V}$}\\ 
\cline{2-7}
&notation&fields&$({\text{SU}}(3)_C,{\text{SU}}(2)_L,
{\text{U}}(1)_Y)$&notation&fields&$
({\text{SU}}(3)_C,{\text{SU}}(2)_L,{\text{U}}(1)_Y$) \\ \hline
&$Q$&($q$, $\sQ$)&(${\bf 3}$, ${\bf 2}$, $\case{1}{6}$)&
$\mrQ$&($\mrq$, $\mrsQ$)&(${\bf \bar{3}}$, ${\bf \bar{2}}$, $-\case{1}{6}$)
\\ \cline{2-7}
matter&$U$&($u$, $\sU$)&(${\bf \bar{3}}$, ${\bf 1}$, $-\case{2}{3}$)&
$\mrU$&($\mru$, $\mrsU$)&(${\bf {3}}$, ${\bf {1}}$, $\case{2}{3}$)
\\ \cline{2-7}
&$D$&($d$, $\sD$)&(${\bf \bar{3}}$, ${\bf 1}$, $\case{1}{3}$)&
$\mrD$&($\mrd$, $\mrsD$)&(${\bf {3}}$, ${\bf {1}}$, $-\case{1}{3}$)
\\\cline{2-7}
multiplets&$L$&($l$, $\sL$)&(${\bf 1}$, ${\bf 2}$, $-\case{1}{2}$)&
$\mrL$&($\mrl$, $\mrsL$)&(${\bf 1}$, ${\bf \bar{2}}$, $\case{1}{2}$)
\\ \cline{2-7}
&$E$&($e$, $\sE$)&(${\bf 1}$, ${\bf 1}$, $1$)&
$\mrE$&($\mre$, $\mrsE$)&(${\bf {1}}$, ${\bf {1}}$, $-1$)\\ \hline
Higgs&$H_1$&$(H_1, \hinod)$&(${\bf 1}$, ${\bf {2}}$, $-\case{1}{2}$)& 
$\mrhd$&$(\mrhd, \mrhinod)$&(${\bf 1}$, ${\bf \bar{2}}$, $\case{1}{2}$)
\\ \cline{2-7}
multiplets&$H_2$&$(H_2, \hinou)$&(${\bf 1}$, ${\bf \bar{2}}$, $\case{1}{2}$)& 
$\mrhu$&$(\mrhu, \mrhinou)$&(${\bf 1}$, ${\bf {2}}$, $-\case{1}{2}$)
\\ \hline
vector& $g$ &$(g,\gluino)$&$({\bf 8},{\bf 1}, 0)$&
$\gadjoint$&$(\gsadjoint,\gfadjoint)$&$({\bf 8},{\bf 1}, 0)$ \\ \cline{2-7}
multiplets&$W$&$(W,\wino)$&$({\bf 1},{\bf 3}, 0)$&
$\wadjoint$&$(\wsadjoint,\wfadjoint)$&$({\bf 1},{\bf 3}, 0)$ \\ \cline{2-7}
&$B$&$(B,\bino)$&$({\bf 1},{\bf 1}, 0)$&
$\badjoint$&$(\bsadjoint,\bfadjoint)$&$({\bf 1},{\bf 1}, 0)$ \\
\end{tabular}
\end{table}
%

For the above particle content, and 
imposing the full $U(2)_{R}$ on the superpotential, 
the theory is scale invariant and is given by
the superpotential (after phase redefinitions) 
\begin{eqnarray}
W/\sqrt{2}
&=&g_3(\mrQ\gadjoint{Q}+\mrU\gadjoint{U}+\mrD\gadjoint{D}) \nonumber \\
&+&g_2(\mrQ\wadjoint{Q}+\mrL\wadjoint{L}+\mrhd\wadjoint{H}_1+
\mrhu\wadjoint{H}_2) \nonumber \\
&+&g_1(\case{1}{6}\mrQ\badjoint{Q}-\case{2}{3}\mrU\badjoint{U}
+\case{1}{3}\mrD\badjoint{D}-\case{1}{2}\mrL\badjoint{L}
+\mrsE\badjoint{E} \nonumber \\
&-&\case{1}{2}\mrhd\badjoint{H}_1+\case{1}{2}\mrhu\badjoint{H}_2).
\label{WN2}
\end{eqnarray} 
After substitution in the Lagrangian (\ref{L}),
the superpotential (\ref{WN2}) gives rise in the usual manner \cite{susy,mssm}
to gauge-quartic and gauge-Yukawa interactions.
All interactions are gauge interactions!
Table~\ref{table:content} and the superpotential~(\ref{WN2}) define
the MN2SSM (in the supersymmetric limit).
%
%


%
\section{Softly broken $N=2$}
\label{sec:soft}
%

Once the MN2SSM is written as an $N = 1$ theory with appropriate spectrum
and global symmetries, as explained above,
supersymmetry breaking translates to the introduction of
$(a)$ SSB dimensionful parameters,
which lift the boson-fermion degeneracy and could also
break the continuous $R$-symmetries,
and of $(b)$ dimensionless parameters
which spoil the constrained $N = 2$ relations between gauge, 
Yukawa and quartic couplings.
We postpone discussion of the latter to Section~\ref{sec:hard},
where we also write all parameters as polynomials in 
the supersymmetry breaking VEVs.
The theory studied in this section is the (global) $N = 2$ SM, the MN2SSM, 
with explicitly and softly broken supersymmetries. 
The breaking is parameterized by
the familiar SSB terms (which also parameterize  supersymmetry 
breaking in $N = 1$ theories).
These terms are soft in the sense that the theory is
at most logarithmically divergent even after their introduction.
The SSB terms  can be chosen to preserve or to break the global symmetries
of the $N = 2$ theory.
However, we leave this issue aside, imposing none of the 
continuous $R$-symmetries on the SSB terms, i.e., 
we assume for now maximal breaking. 
We concentrate instead on $(i)$ those
SSB terms that are unique to $N = 2$ theories and on $(ii)$ those
that break the chiral symmetries (in the SSB scalar potential).
In order to control radiative mixing between the sectors as well
as lepton and baryon number violation we assume that 
the $Z_{2}$ mirror and $R$ parities are conserved, 
unless otherwise specified.
%

In accordance with mirror parity conservation,
the MN2SSM contains in each sector (i.e., the $M_{P}$-even ordinary
and $M_{P}$-odd mirror sectors)
Gaugino mass terms ($M_{\lambda}$), scalar mass terms ($m^{2}_{\phi}$),  
gauge invariant scalar and fermion bilinear terms
($b$, often denoted as $B\mu$, and $\tilde{\mu}$, respectively) 
and trilinear ($A$) terms.
These terms are the the SSB terms familiar from the $N=1$ MSSM,
only in two ``copies''. In addition, trilinear terms can couple
an ordinary particle to two mirror particles.
The $A^{V} y_{i}\sadjoint x_{i}$ 
and ${\cal{A}}^{V} y_{i}\sadjoint x_{j}^{\dagger}$ 
terms in the scalar potential
are an example of such inter-sector couplings.
In addition, a dimensionful mirror parity conserving 
effective superpotential $W = - \mu\hd\hu - \mu^{\prime\prime}\mrhd\mrhu$
may also arise, providing the usual MSSM $\mu$-term and its mirror.
The SSB that may be familiar to the reader from the $N = 1$ MSSM case
are listed in Table~\ref{table:soft1}.
Along side, are listed
their ``mirrored versions'', where in accordance with mirror parity
conservation two of the fields in the operators are substituted 
by their mirror partners. 
If $SU(2)_{R}$ is conserved then the different
SSB (and $\mu$-) parameters are related to each other 
with a significantly smaller number of free parameters.
Note that we included also the non-standard non-holomorphic
trilinear ${\cal A}$-terms and the SSB Higgsino-mass $\tilde{\mu}$-term.
(The SSB Higgsino mass can be absorbed into 
a redefinition of $\mu$ in the superpotential and ${\cal A}$ 
in the SSB scalar potential.)
The next group of SSB operators are those $M_{P}$-even terms 
which are new to the $N=2$ models due to its unique spectrum. 
These are listed listed in Table~\ref{table:soft2}. 
%

\begin{table}
\caption{The $N=1$ MSSM ($R_{P}$ even)
soft supersymmetry breaking terms 
and their $M_{P}$ even mirrored versions.}
\label{table:soft1}
\begin{tabular}{ccc}
&N=1 SSB
&
mirror term\\ \hline
Scalar masses 
&
$m^2_{\sQ}\sQ^{\dagger}\sQ$,  
$m^2_{\sU}\sU^{\dagger}\sU$, 
$m^2_{\sD}\sD^{\dagger}\sD$,
&
${m^2_{\mrsQ}}{{\mrsQ}}^{\dagger}\mrsQ$,  
$m^2_{\mrsU}\mrsU^{\dagger}\mrsU$, 
$m^2_{\mrsD}\mrsD^{\dagger}\mrsD$,\\ 
(Hypermultiplets)
&
$m^2_{\sL}\sL^{\dagger}\sL$, 
$m^2_{\sE}\sE^{\dagger}\sE$,
&
$m^2_{\mrsL}\mrsL^{\dagger}\mrsL$, 
$m^2_{\mrsE}\mrsE^{\dagger}\mrsE$,\\
&
$m^2_{H_1}H_1^{\dagger}{H_1}$, 
$m^2_{H_2}H_2^{\dagger}H_2$&
$m^2_{\mrhd}{\mrhd}^{\dagger}{\mrhd}$, 
$m^2_{\mrhu}{\mrhu}^{\dagger}\mrhu$\\
\\
Gaugino masses
&
$M_1\bino\bino$,  
$M_2\wino\wino$, 
$M_3\gluino\gluino$
&
$M_1^{\prime\prime}\bfadjoint\bfadjoint$,
$M_2^{\prime\prime}\wfadjoint\wfadjoint$, 
$M_3^{\prime\prime}\gfadjoint\gfadjoint$\\
\\
Trilinear operators
&
$A_uH_2\sQ\sU$, 
$A_dH_1\sQ\sD$, 
$A_eH_1\sL\sE$
&
$A_u^{\prime\prime}\hd\mrsQ\mrsU$, 
$A_d^{\prime\prime}\hu\mrsQ\mrsD$, 
$A_e^{\prime\prime}\hu\mrsL\mrsE$\\
\\
&
${\cal A}_u\hd^{\dagger}\sQ\sU$, 
${\cal A}_d\hu^{\dagger}\sQ\sD$, 
${\cal A}_e\hu^{\dagger}\sL\sE$
&
${\cal A}_u^{\prime\prime}\hu^{\dagger}\mrsQ\mrsU$, 
${\cal A}_d^{\prime\prime}\hd^{\dagger}\mrsQ\mrsD$, 
${\cal A}_e^{\prime\prime}\hd^{\dagger}\mrsL\mrsE$\\
\\
Bilinear scalar operators&
$bH_1H_2$&
$b^{\prime\prime}\mrhd\mrhu$\\
\\ 
Bilinear fermion operators&
$\tilde{\mu}\hinod\hinou$&
$\tilde{\mu}^{\prime\prime}\mrhinod\mrhinou$\\ 
\end{tabular}
\end{table}
%

\begin{table}
\label{table:soft2}
\caption{
SSB operators which are unique to the mirror sector in the MN2SSM.} 
\begin{tabular}{cc}
Mirror gauge scalar masses
&
$m^2_{\gsadjoint}\gsadjoint^{\dagger}\gsadjoint$, 
$m^2_{\wsadjoint}\wsadjoint^{\dagger}\wsadjoint$,
$m^2_{\bsadjoint}\bsadjoint^{\dagger}\bsadjoint$\\
\\
Trilinear scalar operators
&
$A^{g}_{Q}\mrsQ\gsadjoint\sQ$, 
$A^{g}_{U}\mrsU\gsadjoint\sU$, 
$A^{g}_{D}\mrsD\gsadjoint\sD$\\
\\
&
$A^{W}_{Q}\mrsQ\wsadjoint\sQ$, 
$A^{W}_{L}\mrsL\wsadjoint\sL$,
$A^{W}_{\hd}\mrhd\wsadjoint\hd$, 
$A^{W}_{\hu}\mrhu\wsadjoint\hu$\\
\\
&
$A^{B}_{Q}\mrsQ\bsadjoint\sQ$, 
$A^{B}_{U}\mrsU\bsadjoint\sU$, 
$A^{B}_{D}\mrsD\bsadjoint\sD$,  
$A^{B}_{l}\mrsL\bsadjoint\sU$, 
$A^{B}_{E}\mrsE\bsadjoint\sE$\\
&
$A^{B}_{\hd}\mrhd\bsadjoint\hd$, 
$A^{B}_{\hu}\mrhu\bsadjoint\hu$\\
\\
&
${\cal{A}}^{W}_{H}\mrhd\wsadjoint\hu^{\dagger}$, 
${\cal{A}}^{B}_{H}\mrhd\bsadjoint\hu^{\dagger}$
\end{tabular}
\end{table}
%

If mirror parity and $U(1)_{R}^{N=2}$
are not conserved (see Section~\ref{sec:mixing}) 
the effective superpotential could contain $W = \mu^{\prime}XY$ terms 
with $\mu^{\prime}$ being an arbitrary mass parameter. In addition,
mirror parity violating (MPV)
SSB terms can also mix the two sectors. 
For completeness, we list the MPV SSB operators in Table~\ref{table:soft3}.
The mixing terms
$q\mrq$, $u\mru$ for the third family can
play an important role in generating the heavy top mass.
Similarly, $l\mrl$ mixing can play a role in the
generation of light neutrino masses. 
This will be discussed in detail in sections~\ref{sec:mixing}
and ~\ref{sec:neutrino}, respectively. Otherwise, such terms are
assumed to be absent. This completes the listing of (dimensionful)
SSB terms in the MN2SSM.
%

\begin{table}
\caption{Mirror parity violating SSB operators in the MN2SSM.}
\label{table:soft3}
\begin{tabular} {cc}
Trilinear operators
&
$A_u^{\prime}\mrhd\sQ\sU$, 
$A_d^{\prime}\mrhu\sQ\sD$, 
$A_e^{\prime}\mrhu\sL\sE$\\
\\
&
$A_u^{\prime\prime\prime}\mrhu\mrsQ\mrsU$, 
$A_d^{\prime\prime\prime}\mrhd\mrsQ\mrsD$, 
$A_e^{\prime\prime\prime}\mrhd\mrsL\mrsE$\\
\\
&
${\cal A}_u^{\prime}\mrhu^{\dagger}\sQ\sU$, 
${\cal A}_d^{\prime}\mrhd^{\dagger}\sQ\sD$, 
${\cal A}_e^{\prime}\mrhd^{\dagger}\sL\sE$\\
\\
&
${\cal A}_u^{\prime\prime\prime}\mrhd^{\dagger}\mrsQ\mrsU$, 
${\cal A}_d^{\prime\prime\prime}\mrhu^{\dagger}\mrsQ\mrsD$, 
${\cal A}_e^{\prime\prime\prime}\mrhu^{\dagger}\mrsL\mrsE$\\
\\
Scalar mixing
&
$b_{\hd}^{\prime}H_1\mrhd$,  
$b_{\hu}^{\prime}H_2\mrhu$,
$b_{Q}^{\prime}\sQ\mrsQ$, 
$b_{U}^{\prime}\sU\mrsU$, 
$b_{D}^{\prime}\sD\mrsD$, 
$b_{L}^{\prime}\sL\mrsL$, 
$b_{E}^{\prime}\sE\mrsE$\\
\\
Chiral fermion mixing
&
$\tilde{\mu}_{\hd}^{\prime}\hinod\mrhinod$,  
$\tilde{\mu}_{\hd}^{\prime}\hinou\mrhinou$,
$\tilde{\mu}_{q}^{\prime}q\mrq$, 
$\tilde{\mu}_{u}^{\prime}u\mru$, 
$\tilde{\mu}_{d}^{\prime}d\mrd$, 
$\tilde{\mu}_{l}^{\prime}l\mrl$, 
$\tilde{\mu}_{e}^{\prime}e\mre$\\ 
\\ 
Gauge fermion mixing
&
$M_{3}^{\prime}\gluino\gfadjoint$, 
$M_{2}^{\prime}\wino\wfadjoint$, 
$M_{1}^{\prime}\bino\bfadjoint$\\

\end{tabular}
\end{table}
%

The softly broken MN2SSM resembles, not surprisingly, 
an extended MSSM. For example, 
consider electroweak symmetry breaking (EWSB).
In the $N=1$ MSSM EWSB is triggered by the SSB-terms in the Higgs potential.
EWSB in the MN2SSM can be induced, in general,  
by any combination of the four Higgs doublets and the triplet $\phi_{W}$.
However, mirror parity conservation allows only the ordinary MSSM
Higgs doublets $\hd$ and $\hu$ to acquire VEVs.
(Independently of the parity considerations,
a triplet VEV is strongly constrained by
electroweak data and has to practically vanish \cite{pdg}.)
From the discussion of the effective Yukawa couplings
it will become evident that in fact it is sufficient
that only one Higgs doublet acquires a VEV, which can then be truly
identified with the SM Higgs boson. However, here we assume, for simplicity,
that the MSSM realization of EWSB with two Higgs doublets is reproduced
with the usual 
Higgs doublets $\hd$ and $\hu$ receiving non-zero VEVs 
$v_1$ and $v_2$, respectively. This is achieved  by  
adjusting the soft parameters which enter the Higgs potential
such that $(m_{\hd}^{2} + \mu^{2})(m_{\hu}^2 +\mu^{2}) < |b|^{2}$.
Though introduced here by hand, this relation could be satisfied
via a generalization of the MSSM radiative 
symmetry breaking mechanism \cite{mssm}.
However, since we do not discuss any specific pattern
of the boundary conditions to the SSB parameters, we postpone
discussion of their renormalization for future works.
Defining, as usual, $\tan\beta=v_2/v_1$, 
the $Z$ boson mass $m_Z$ is then given by
\begin{equation}
\frac{1 + \Delta_{\rm hard}}{2}
m_{Z}^2=\frac{m_{H_{1}}^2-m_{H_{2}}^2\tan^2\beta}{\tan^2\beta-1}
-\mu^2,
\label{minimization}
\end{equation}
where $\Delta_{\rm hard}$ contains the effects of
hard-supersymmetry corrections to the quartic terms in 
Higgs potential, which are discussed in Section~\ref{sec:hard}.
In the softly broken MN2SSM $\Delta_{\rm hard} \equiv 0$.  
%

There is, however, an important difference between the MSSM and the MN2SSM.
While the gauge bosons get masses via the usual Higgs mechanism
and (\ref{minimization}) reduces to its MSSM form, 
the softly broken MN2SSM contain no mass terms
for the usual and mirror fermions. This is due to the absence of 
tree-level Higgs Yukawa couplings.   
Is this naive MSSM-like softly broken MN2SSM  
in which supersymmetry is broken only by dimensionful parameters
then viable? The key to the answer lies with
the trilinear terms, which are the only terms that break the chiral symmetries
in the scalar potential and can therefore induce fermion
masses at the quantum level, $m_{f} \propto Am_{\lambda},\,
A^{\prime\prime}m_{\psi_{V}}$ 
(where the gaugino and mirror gaugino masses are 
responsible for the breaking of fermion number), and similarly for the mirror
fermions. The viability of this mechanism
is discussed in the next section.
%
%
%


%
\section{Radiative fermion masses}
\label{sec:loop}
%

We begin the discussion
of fermion mass generation with the discussion of radiative fermion masses.
(More general mechanisms will be discussed in Section~\ref{sec:tree},
following the generalization of the softly broken MN2SSM in 
Section~\ref{sec:hard}.)
The observation that chiral symmetries could be primarily broken
in the scalar potential and that the 
fermion spectrum in supersymmetry could
arise radiatively was first made in the context of $N =2$ supersymmetry
\cite{west}, though it was studied most extensively in the case of $N = 1$
supersymmetry\cite{BFPT}. 
It provides an avenue for the generation of fermion
masses in the softly broken MN2SSM which was discussed
in the previous section. Such a mass generation mechanism has the advantage
that it could be accommodated in any scenario of SSB which
includes the generation of trilinear terms. On the other hand, it
is highly constrained. 
%

\begin{figure}
\PSbox{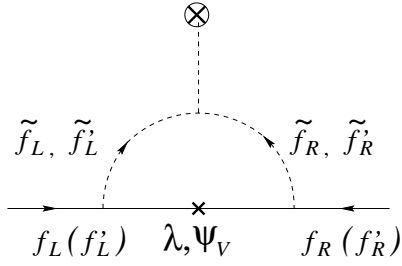 hoffset=160 voffset=20 hscale=50
vscale=50}{6.0in}{1.5 in}
\caption{One-loop contribution to fermion mass from soft chiral symmetry
 breaking.}
\label{fig:radiative}
\end{figure}
%

At one loop, the extended supersymmetry
gauge interactions lead to loops such as in Fig.~\ref{fig:radiative},
with external ordinary fermions
where the sfermion $\tilde{f}$ and gaugino $\lambda$
(or mirror sfermion $\tilde{f}^{\prime}$ and mirror gaugino $\psi_{V}$)
propagate in the loop. Equivalent loops exist with external mirror fermions.
The sfermion left-right mixing terms $m^{2}_{\chi SB}$
proportional to $
A\langle H\rangle$, where $A$ here is a chiral symmetry
breaking trilinear parameter
and $\langle H \rangle$ is a $SU(2)_{L}$-breaking Higgs (doublet) VEV, 
generates a finite 
contribution to a chiral fermion mass,
\begin{eqnarray}
 -m_f &=&   
\frac{\alpha_s}{2 \pi} \,C_f  \, \left[ 
m^{2}_{\chi SB}M_{3}
\,I(m^{2}_{\tilde{f}_1}, m^{2}_{\tilde{f}_2}, M^2_{3})
+ m^{2\,\prime\prime}_{\chi SB}M_{3}^{\prime\prime}
\,I(m^{2}_{\tilde{f}_1^{\prime}}, m^{2}_{\tilde{f}_2^{\prime}}, 
{M^{\prime\prime}_{3}}^{2}) 
\right] 
\nonumber \\
&&       
+\frac{\alpha^\prime}{2\pi}Y_{f_{L}}Y_{f_{R}} 
\left[
m^{2}_{\chi SB}M_{1} 
\,I(m^{2}_{\tilde{f}_1}, m^{2}_{\tilde{f}_2}, M^{2}_{1})  
+m^{2\,\prime\prime}_{\chi SB}M_{1}^{\prime\prime} 
\,I(m^{2}_{\tilde{f}_1}, m^{2}_{\tilde{f}_2}, {M^{\prime\prime}_{1}}^{2})  
\right]
\,,
\label{loopmass}
\end{eqnarray}
which generalizes the expressions given in Ref.~\cite{BFPT}.
($m_{\tilde{f}_{i}}^{2}$, $m_{\tilde{f}_{i}^{\prime}}^{2}$
are the sfermion and mirror sfermion mass eigenvalues and
$m^{2\,\prime\prime}_{\chi SB} = A^{\prime\prime}\langle H \rangle$.)
%

The first and second terms in Eq.~(\ref{loopmass})
correspond to the strong (gluino and mirror gluino) 
and hypercharge (the Bino neutralino and its mirror)
contributions, respectively, where $\alpha_{s}$ 
and $\alpha^{\prime}$ are the strong and hypercharge
couplings, $C_{f} = 4/3, 0$ for quarks and leptons, respectively,
and $Y_{f}$ is the fermion hypercharge.
We assume, for simplicity,  no neutralino mixing. 
The function $I$ is the loop function
\begin{equation}
 I(m_1^2,m_2^2,m_\psi^2)  =   - \frac{
 m_1^2 \,m_2^2 {\rm ln} ( m_1^2/ m_2^2 ) +
 m_2^2 \,m_\psi^2 {\rm ln} ( m_2^2/ m_\psi^2 )  +
 m_\psi^2 \,m_1^2 {\rm ln} ( m_\psi^2/ m_1^2 )
                                         }
{(m_1^2 -m_2^2)(m_2^2 -m_\psi^2)(m_\psi^2 -m_1^2)
          }\,.
\label{Ifunc}
\end{equation}
which typically behaves as 
$\sim {\cal{O}}(1/{\rm max}(m_1^2, m_2^2,m_\psi^2))$.
The dependence on the left--right
squark or slepton mixing,
$m_{\chi SB}^2 = A \langle H_\alpha \rangle$ and its mirror
$m_{\chi SB}^{2\,\prime\prime} = A^{\prime\prime} \langle H_\alpha \rangle$,
and on the chiral violation arising from the gaugino
and mirror-gaugino Majorana masses,
$M_{3},\,M_{3}^{\prime\prime}$, and/or 
$M_{1},\,M_{1}^{\prime\prime}$, is 
explicitly displayed in Eq.~(\ref{loopmass}).
%

By observation, the natural size of the resulting
fermion mass is \begin{equation}
m_{f} \sim \left\{\begin{array}{cc}
\frac{\alpha^{\prime}}{4\pi}\langle H \rangle 
\lesssim {\cal{O}}(100\,\,{\rm MeV})&
{\rm Lepton}\\
&\\
&\\
\frac{2\alpha_{s}}{3\pi}\langle H \rangle
\lesssim {\cal{O}}(1\,\,{\mbox{\rm GeV}})&
{\rm Quark}
\end{array}\right.,
\label{loopmass2}
\end{equation}
and similarly for the mirror fermions.
This is the appropriate mass range for most of the ordinary fermions,
but not for the mirror fermions (and $t$-quark), whose masses are given 
by a similar expression.
In the approximation Eq.~(\ref{loopmass2}) it was implicitly assumed that
all of the SSB parameters are of the same order of magnitude,
e.g., $m_{\chi SB}^{2}m_{\lambda}
I(m^{2}_{\tilde{f}_1}, m^{2}_{\tilde{f}_2}, m^{2}_{\lambda}) \simeq 
\langle H \rangle \times {\cal{O}}(1)$,  
and therefore cannot change the order of magnitude of the 
resulting fermion mass.
%

More generally, however, the fermion mass can grow as
$|A/{\rm max}(m_{\tilde{f}},\,m_{\lambda})|$ or
$|A/{\rm max}(m_{\tilde{f}^{\prime}},\,m_{\psi_{V}})|$.
Thus, it may appear that the fermion mass could in fact be as large as 
${\cal{O}}(\langle H \rangle)$, which is  the correct mass range
for the mirror fermions,
provided that the size of the
trilinear coupling roughly equals in size to the inverse of a loop factor.
Unfortunately, this cannot be the case. As already noted in Ref.~\cite{west},
in the relevant limit of no tree-level
Yukawa couplings the  trilinear parameters destabilize
the scalar potential (along the equal field direction)
leading to color and charge breaking. In particular, 
the trilinear couplings cannot be too large. 
It was recently noted \cite{BFPT}
that the scalar potential may be stabilized if 
effective quartic coupling are generated by the  decoupling of  
chiral superfields with masses of the order of the SSB parameters
or in the presence of non-holomorphic trilinear terms 
${\cal{A}}\hd^{\dagger}\sQ\sU$ etc.
(which do not correspond to flat directions of the scalar potential).
Even though the potential is stable in this case, requiring
color and charge conservation in the global (or meta-stable) minimum still
constrains $A/m_{\tilde{f}}$ from above. (See Fig.~7 of Ref.~\cite{BFPT}.) 
In the next section we will show that in theories with low-energy
supersymmetry breaking there could also appear arbitrary 
and hard supersymmetry breaking quartic couplings $\kappa$, 
which could further stabilize the potential. 
Nevertheless, given the stability constraints 
$\kappa \gtrsim \sqrt{3}A/m_{\tilde{f}}$
\cite{destabilize,BFPT}, then $A$ cannot not be 
sufficiently large to accommodate the heavy mirror fermions
(for perturbative values of the quartic couplings).
%

We conclude that radiative fermion mass generation
leads to a viable scenario only in the case of (most of) 
the ordinary fermions. In particular,
even though the gauge loop can break the ordinary-mirror mass degeneracy,
provided that it is already broken by the SSB parameters, 
it cannot provide the required two or more orders of magnitude separation
between the ordinary and mirror fermion spectra.
As for the SM fermions,
in the limit that the gaugino-sfermion loop dominates
(or equivalently only its mirror loop dominates)
this reduces to the case studied in Ref.~\cite{BFPT}, 
with distinctive phenomenology
and signatures. If both loop classes contribute, the flavor structure
becomes more complicated, but the model still maintains 
the same essential features and signatures. We refer the interested reader
to Ref.~\cite{BFPT} for an extensive discussion.
%

We note in passing that even if right-handed neutrinos are introduced,
the SM neutrinos would remain massless if the fermion spectrum
is indeed generated via such (supersymmetric) gauge loops,
since the right-handed neutrino is a gauge singlet in the SM.
(We return to the neutrinos in Section~\ref{sec:neutrino}.)
The lightness of the neutrinos could be explained in this context
by extending the SM by an extremely weakly coupled Abelian factor
under which the right-handed neutrinos are not singlets.
%

Radiative mass generation for the SM fermions must go hand in hand
with a mechanism to lift the ordinary-mirror matter mass degeneracy
(and with a mechanism to lift the $t$-quark mass).
Such mechanisms will be studied in the following sections,
and require one to consider a more general 
parameterization of supersymmetry breaking. We turn to a general
classification of supersymmetry breaking operators in the next section.
%
%


%
\section{Classification of supersymmetry-breaking operators}
\label{sec:hard}
%

In Section~\ref{sec:loop},
the possibility of radiatively induced Yukawa couplings
in the effective low-energy theory was shown
to lead, regardless of its details, to a typical mass range of 
$m_{q},\,m_{q^{\prime}} \lesssim (\alpha_{s}/\pi)\langle H \rangle$ 
for the ordinary and mirror quarks, and 
$m_{l},\,m_{l^{\prime}} \lesssim (\alpha^{\prime}/\pi)\langle H \rangle$ for 
the ordinary and mirror leptons. 
This is a natural and sufficient solution in the case of most of 
the ordinary fermions
(we postpone that discussion of the third family fermions to
Section~\ref{sec:mixing}) but not in the case their mirrors. 
The mirror fermion spectrum is constrained by experiment, 
$m_{f^{\prime}} \simeq \langle H \rangle$, which implies tree-level 
effective Yukawa couplings.  
(The constraints, however, are model dependent.)
Typically, a Kahler function $K$ describing 
the effective low-energy theory includes 
tree-level non-renormalizable Yukawa operators, either supersymmetry
conserving, soft, or hard supersymmetry breaking. 
In order to consider such operators
one must step out of the softly broken MN2SSB framework
of Section~\ref{sec:soft} and systematically include all relevant 
supersymmetry breaking operators.
This will be done in this section. Concrete realizations
of tree-level Yukawa couplings in the MN2SSM will be considered in the
next section.
%

Our classification applies
to any theory which can effectively be described by $N=1$ superfields,
and will be adopted to the $N=2$ case only in the next section.
In typical $N = 1$ (high-energy) supergravity model building with
supersymmetry breaking scale $F \simeq \mweak\mplanck$,
the Yukawa (and quartic) operators listed below are 
proportional to $(\mweak/\mplanck)^{n}$, $n=1,\,2$, 
and hence are often omitted.
(Nevertheless, even in that case such terms can shift any boundary conditions
for the SSB by ${\cal{O}}(100\%)$ \cite{boundary}.) 
This proportionality, however, cannot hold in the $N = 2$ case if 
it is to be phenomenologically viable. 
Requiring a viable phenomenology
constrains the size and symmetries of the effective Yukawa couplings, 
and hence the scales and symmetries
of the $N = 2 $ theory and of the theory below the 
supersymmetry breaking scale. 
Indeed, since the gravitino masses
are somewhat arbitrary in the $N=2$ case \cite{GPZ}, there is no reason 
to impose any relation analogous to the above $N=1$ supergravity relation,
even when gravity is introduced. 
After considerations of all operators
we will set the supersymmetry breaking scale simply
by requiring sizeable tree-level Yukawa couplings and 
stability of the theory against hard operators, and compare it
to the requirement that $N=2$ supersymmetry plays a role
in the resolution of the hierarchy problem.
(We will find the two requirements to be consistent.)
%

In Secs.~\ref{sec:soft} and \ref{sec:loop} 
we adopted the description of supersymmetry 
breaking in terms of an explicitly
but softly broken global $N=1$ theory with a second supersymmetry
only implicitly manifest in the symmetries of the (super)potential,
and its breaking corresponding to explicit breaking of the relevant symmetries
in the SSB potential.
Though we are about to extend and generalize this description
to include dimensionless  $N=2$ and $N=1$ breaking 
couplings, the same modular description of supersymmetry breaking will
prove to be a powerful classification tool here as well.
Its generalization corresponds to the replacement of explicit soft breaking
terms with the spurion formalism \cite{GG}.
A spurion field $X = \theta^{2}F_{X}$
parameterizes the manifest $N=1$ breaking, and non-renormalizable
operators which couple $X$ to the MN2SSM fields parameterize 
the explicit $N=2$ (exchange symmetry) breaking.
The non-renormalizable operators scale as inverse powers
of the $N=2$ breaking scale $M$. 
The convenient spurion formalism is available only in this $N=1$ formulation.
(Note that the spurion $X$ is 
not to be confused with the generic $X$ superfield 
component of a hypermultiplet in Section~\ref{sec:theory}.)
%

Indeed, one could arbitrarily write
down in the infra-red theory Yukawa and quartic couplings 
whose presence leads to quadratically
divergent quantum correction to various two-point functions,
and which are therefore said to be hard supersymmetry breaking. However, if
one requires that in certain $M \rightarrow \infty$ the full (global) $N=2$
supersymmetry is recovered, then these couplings must fall into certain
categories of non-renormalizable operators.
It is further reasonable to assume that the Kahler potential 
(which is not protected by non-renormalization theorems)
rather than the superpotential accommodates these operators.\footnote{
These operators are induced, in principle, by the dynamics at the scale $M$.
The resulting low-energy effective Kahler potential is not derived, in general,
from a holomorphic prepotential function $P$,
$K(\Phi) \neq {\rm Im}\left[\Phi^{\dagger}(\partial P /\partial \Phi)\right]$.}
(As we shall see below, both possibilities of Kahler or superpotential 
operators lead in practice to couplings of the same size.)
The non-renormalizable Kahler potential operators which link
the spurion and the MN2SSM fields
do not preserve the global symmetries of the full
$N=2$ theory, which is equivalent to the symmetry violations
by the SSB potential in the previous sections.
In addition, non-vanishing values of $X = \theta^{2}{F}_{X}$ parameterize the  
breaking of the manifest $N=1$ supersymmetry
as well as replace the non-renormalizable operators with explicit
($N=2$ and $N=1$) supersymmetry breaking terms in the low-energy potential.
Note that the spurion and its mirror $(X, X^{\prime})$ transform 
as a doublet under the $SU(2)_{R}$ exchange $R$-symmetry, which
implies that a non-vanishing VEV $\langle F_{X} \rangle$ automatically breaks
it. (The $SU(2)_{R}$ of $N=2$ allows one to rotate the supersymmetry
breaking VEV such that the mirror $F_{X^{\prime}} = 0$.)
This parameterization has two breaking parameters, $F_{X}$ and $M$,
corresponding to the spurion VEV and (inverse) couplings, respectively. 
It corresponds to an one-step breaking scenario, $N=2 \rightarrow N=0$, 
for $F_{X} \simeq M^{2}$, which we will assume.
%

We now turn to a general classification of $K$ operators.
We do not impose any of the global symmetries which parameterize
the second supersymmetry, a subset of which can  survive its breaking.
(This will be done in the next section.) 
The effective low-energy Kahler potential of a rigid 
$N = 1$ supersymmetry theory is given by
\begin{eqnarray}
K &=&  K_{0}(X,X^{\dagger}) + K_{0}(\Phi,\Phi^{\dagger}) 
\nonumber \\
&+& \frac{1}{M}K_{1}(X,X^{\dagger},\Phi,\Phi^{\dagger})
+ \frac{1}{M^{2}}K_{2}(X,X^{\dagger},\Phi,\Phi^{\dagger})
\nonumber \\
&+& \frac{1}{M^{3}}K_{3}(X,X^{\dagger},\Phi,\Phi^{\dagger},D_{\alpha},W_{\alpha})
+ \frac{1}{M^{4}}K_{4}(X,X^{\dagger},\Phi,\Phi^{\dagger},D_{\alpha},W_{\alpha})
+ \cdots 
\label{Kgeneral}
\end{eqnarray}
where $X$ is the spurion and $\Phi$ are the (ordinary and mirror) chiral 
superfields of the low-energy MN2SSM theory. 
$D_{\alpha}$ is the covariant derivative with respect 
to the (explicit) superspace chiral coordinate $\theta_{\alpha}$, 
and $W_{\alpha}$ is the $N = 1$ gauge supermultiplet in its chiral 
representation, $W_{\alpha} \sim \lambda_{\alpha} + \theta_{\alpha} V$.
Once a separation between supersymmetry breaking field $X$ and low-energy
$\Phi$ fields is imposed, there is no tree-level renormalizable
interaction between the two sets of fields, and their mixing
can arise only at the non-renormalizable level $K_{l \geq 1}$.
(This separation is quite natural in the context of $N =2$
if $X$ and $\Phi$ transform under different gauge groups,
in particular if $X$ is a gauge singlet field.)
%

The superspace integration  ${\cal{L}}_{D} = 
\int d^{2}\theta d^{2}\bar{\theta}K$
reduces $K_{1}$ and $K_{2}$ to the usual SSB terms, as well as the 
superpotential $\mu$-parameter $W \sim \mu\Phi^{2}$, which were 
discussed in the previous section. It also contains Yukawa operators
$W \sim  y\Phi^{3}$ which can appear in the effective low-energy 
superpotential. These are summarized
in Tables~ \ref{table:o1} and~\ref{table:o1b}.
(We did not include linear terms that may appear
in the case of a singlet superfield.) Finally, The last 
term in Table~\ref{table:o1} contains correlated but unusual quartic 
and Yukawa couplings. 
They are soft as they involve at most logarithmic divergences.
%

Integration over $K_{3}$ produces the non-standard soft terms, also
discussed in the previous section. 
These are summarized in Table~\ref{table:o2}.
It also generates contributions to the (``standard'') 
$A$ and gaugino-mass terms.
These terms could arise at lower orders in $\sqrt{F}/M$ from integration 
over holomorphic functions 
(and in the case of $A$, also from $K_{1}$). 
However, this  is equivalent to integration 
over $K$ if 
$\int d^{2}\bar{\theta}(X^{\dagger}/M^{2}) \simeq 1$.  
Note that in the presence of superpotential Yukawa couplings, a 
Higgsino $\tilde{\mu}$
term can be rotated to a combination of $\mu$ and ${\cal{A}}$ terms
and vice versa.  The two terms, however, are not necessarily equivalent in 
our case since $N=2$ forbids chiral superpotential Yukawa couplings.
In the case of the mirror gaugino $\fadjoint$, our MN2SSM notation
replaces $\tilde{\mu}$ with $M^{\prime\prime}/2$.
%

Lastly, superspace integration over $K_{4}$ leads to dimensionless
hard operators. These are summarized in Table~\ref{table:o3}.
(Hard  operators were also summarized recently in Ref.~\cite{martin}.)
Higher orders in $(1/M)$ can be safely neglected
as supersymmetry and the superspace integration allow only
a finite expansion in $\sqrt{F_X}/M$, that is ${\cal{L}}
= f[F^{n}_X/M^{l}]$ with $n \leq 2$ and
$l$ is the index $K_{l}$ in expansion Eq.~(\ref{Kgeneral}). 
Hence, terms with $l > 4$ are suppressed by at least 
$(\langle X \rangle /M)^{l-4}$.
We will assume the limit $\langle X \rangle \ll M$
for the $N=1$ supersymmetry preserving VEV  $\langle X \rangle$, i.e., 
$X \sim \theta^{2}F_{X}$, 
so that all such operators can indeed be neglected
and the expansion is rendered finite.
%

\begin{table}
\caption{The soft supersymmetry breaking terms as operators
contained in $K_{1}$ and $K_{2}$. $\Phi = \phi + \theta\psi + \theta^{2}F$
is a low-energy superfield while $X$, $\langle F_{X} \rangle  \neq 0$, 
parameterizes supersymmetry breaking. 
$F^{\dagger} = \partial W/ \partial \Phi$.}
\label{table:o1}
\begin{tabular}{cc}
ultra-violet $K$ operator & infra-red ${\cal{L}}_{D}$ operator  \\ \hline
&\\
$\frac{X}{M}\Phi\Phi^{\dagger} + \hc$ & $A \phi F^{\dagger}+ \hc$ \\
&\\
$\frac{XX^{\dagger}}{M^{2}}\Phi\Phi^{\dagger}$ + \hc& 
$\frac{m^{2}}{2}\phi\phi^{\dagger} + \hc $ \\
&\\
$\frac{XX^{\dagger}}{M^{2}}\Phi\Phi + \hc $ & $b\phi\phi + \hc$ \\
&\\
$\frac{X^{\dagger}}{M^{2}}\Phi^{2}\Phi^{\dagger}
+ \hc $ & $\kappa\phi^{\dagger}\phi F + \hc$ \\
 & $y\phi^{\dagger}\psi\psi + \hc$ \\

\end{tabular}
\end{table}
%

\begin{table}
\caption{
The effective renormalizable
$N = 1$ superpotential $W$ operators
contained in $K_{1}$ and $K_{2}$, ${\cal{L}} = \int d^{2}\theta W$. 
Symbols are defined in Table \ref{table:o1}.
}
\label{table:o1b}
\begin{tabular}{cc}
ultra-violet $K$ operator & infra-red $W$ operator  \\ \hline
&\\
$\frac{X^{\dagger}}{M}\Phi^{2}+ \hc$ & $\mu\Phi^{2}$ \\
&\\
$\frac{X^{\dagger}}{M^{2}}\Phi^{3}+ \hc$ & $y\Phi^{3}$ \\

\end{tabular}
\end{table}
%

\begin{table}
\caption{The non-standard or semi-hard  
supersymmetry breaking terms as operators
contained in $K_{3}$. 
$W^{\alpha}$ is the $N=1$ chiral
representation of the gauge supermultiplet and $\lambda$ is the respective
gaugino. $D_{\alpha}$ is the covariant derivative with respect to the 
(explicit) superspace coordinate $\theta_{\alpha}$.
All other symbols are as in Table \ref{table:o1}}
\label{table:o2}
\begin{tabular}{cc}
ultra-violet $K$ operator & infra-red ${\cal{L}}_{D}$ operator  \\ \hline
&\\
$\frac{XX^{\dagger}}{M^{3}}\Phi^{3} + \hc$ 
& $A\phi^{3} +\hc$ \\
&\\
$\frac{XX^{\dagger}}{M^{3}}\Phi^{2}\Phi^{\dagger} + \hc$ 
& ${\cal{A}}\phi^{2}\phi^{\dagger} +\hc$ \\
&\\
$\frac{XX^{\dagger}}{M^{3}}D^{\alpha}\Phi D_{\alpha}\Phi + \hc $ & 
$\tilde{\mu}\psi\psi + \hc$\\ 
&\\
$\frac{XX^{\dagger}}{M^{3}}D^{\alpha}\Phi W_{\alpha} + \hc $ & 
$M^{\prime}_{\lambda}\psi\lambda + \hc$ \\ 
&\\
$\frac{XX^{\dagger}}{M^{3}}W^{\alpha} W_{\alpha} + \hc $ & 
$\frac{M_{\lambda}}{2}\lambda\lambda + \hc$ \\ 

\end{tabular}
\end{table}
%

\begin{table}
\caption{The dimensionless hard  
supersymmetry breaking terms as operators
contained in $K_{4}$. 
Symbols are defined as in Tables \ref{table:o1} and \ref{table:o2}.
}
\label{table:o3}
\begin{tabular}{cc}
ultra-violet $K$ operator & infra-red ${\cal{L}}_{D}$ operator  \\ \hline
&\\
$\frac{XX^{\dagger}}{M^{4}}\Phi D^{\alpha}\Phi D_{\alpha} \Phi + \hc$ 
& $y\phi\psi\psi  +\hc$ \\
&\\
$\frac{XX^{\dagger}}{M^{4}}\Phi^{\dagger}D^{\alpha}\Phi D_{\alpha} \Phi + \hc$ 
& $y\phi^{\dagger}\psi\psi  +\hc$ \\
&\\
$\frac{XX^{\dagger}}{M^{4}}\Phi D^{\alpha}\Phi W_{\alpha}  + \hc$ 
& $\bar{y}\phi\psi\lambda  +\hc$ \\
&\\
$\frac{XX^{\dagger}}{M^{4}}\Phi^{\dagger}D^{\alpha}\Phi W_{\alpha} + \hc$ 
& $\bar{y}\phi^{\dagger}\psi\lambda  +\hc$ \\
&\\
$\frac{XX^{\dagger}}{M^{4}}\Phi W^{\alpha} W_{\alpha} + \hc$ 
& $\bar{y}\phi\lambda\lambda  +\hc$ \\
&\\
$\frac{XX^{\dagger}}{M^{4}}\Phi^{\dagger}W^{\alpha} W_{\alpha} + \hc$ 
& $\bar{y}\phi^{\dagger}\lambda\lambda  +\hc$ \\
&\\
$\frac{XX^{\dagger}}{M^{4}}\Phi^{2}\Phi^{\dagger\, 2} + \hc$ 
& $\kappa(\phi\phi^{\dagger})^{2} +\hc$ \\
&\\
$\frac{XX^{\dagger}}{M^{4}}\Phi^{3}\Phi^{\dagger} + \hc$ 
& $\kappa\phi^{3}\phi^{\dagger} +\hc$ \\

\end{tabular}
\end{table}
%

It is useful for our purposes to identify those terms in $K$ 
which can break the chiral symmetries
and generate the desired Yukawa terms in the low-energy effective theory.
Clearly, the relevant terms in tables~\ref{table:o1} and~\ref{table:o2} 
can be identified with the chiral symmetry breaking $A$- and ${\cal {A}}$-terms
(with any number of primes)
which couple the matter sfermions
to the Higgs fields of electroweak symmetry breaking
and which were discussed in the previous section.  
Note that since in $N=2$ there are no chiral
terms in the superpotential then chiral-symmetry breaking
$A$-terms can only arise from $K_{3}$.
More importantly, and as advertised above,
a generic Kahler potential is also found to contain 
tree-level chiral Yukawa couplings.
These include ${\cal{O}}(F_X/M^{2})$ supersymmetry conserving and soft
couplings and ${\cal{O}}(F^{2}_X/M^{4})$ hard chiral symmetry breaking 
couplings. The relative importance and the potentially destabilizing properties
of the different operators must be addressed
before any symmetry-derived selection rules are applied. 
Both issues point to the more fundamental questions that one needs to address:
What are the scales $\sqrt{F_X}$ and $M$ and what is their
relation to the cut-off scale $\Lambda$. 
%

We have $\sqrt{F_{X}} \simeq M \simeq {\cal{O}}({\rm TeV})$ 
from the requirement that $N=2$ supersymmetry plays a role 
in the solution of the SM hierarchy problem.
In addition, the cut-off scale for any such calculation is the scale
of $N=2$ restoration above which $F_{X} = 0$, i.e., $\Lambda \simeq M$.
In this case, all of the dimensionful couplings could be 
in principle $\sim{\cal{O}}(1)$,
regardless of their softness or order in $F_{X}/M^{2}$.
This is desired for the Yukawa couplings of the mirror fermions.
It is important to note, however, that quartic couplings are also large.
(We mentioned the latter effect in the previous section.)
One has to confirm that this choice is not destabilized
when the hard operators, which are large, are included. 
In order to do so, consider the implication of the hardness of the operators
contained in $K_{4}$. Yukawa and quartic couplings
can destabilize the scalar potential by corrections to the mass terms
$\Delta m^{2}$ of the order of
\begin{equation}
\Delta m^{2} \sim \left\{\begin{array}{c}
\frac{\kappa}{16\pi^{2}}\Lambda^{2} \sim 
\frac{1}{16\pi^{2}}\frac{F^{2}_{X}}{M^{4}}\Lambda^{2} \sim 
\frac{1}{16\pi^{2}}\frac{F^{2}_{X}}{M^{2}} \sim 
\frac{1}{16\pi^{2}c_{m}}m^{2}
\\
\\
\frac{y^{2}}{16\pi^{2}}\Lambda^{2} \sim 
\frac{1}{16\pi^{2}}\frac{F^{4}_{X}}{M^{8}}\Lambda^{2} \sim 
\frac{1}{16\pi^{2}}\frac{F^{4}_{X}}{M^{6}} \sim 
\frac{1}{16\pi^{2}c_{m}}m^{2}\frac{m^{2}}{M^{2}},
\end{array} \right.
\label{QD}
\end{equation}
where we identified $\Lambda \simeq M$ and $c_{m}$ is a dimensionless
coefficient omitted in Table~\ref{table:o1}, $m^{2}/2 = c_{m}F_{X}^{2}/M^{2}$.
The hard operators were substituted by the appropriate
powers of $F_{X}/M^{2}$ (and are $\sim{\cal{O}}(1)$). 
Once $M$ is identified as the cut-off scale above which
the the full supersymmetry is restored,
then these terms are harmless as the contributions are bound
from above by the tree-level  scalar  squared-mass parameters. 
%

This observation is valid for the $N = 1$ case
whether it is constrained by the $N=2$ symmetries
or not (and extends to the case of non-standard soft operators
in the presence of a singlet). 
In fact, such hard divergent corrections are well known in  $N=1$ supergravity
with $\Lambda = M = \mplanck$, where they perturb
any given set of tree-level boundary conditions for the SSB 
parameters \cite{boundary}. 
In theories with low-energy supersymmetry breaking
$\sqrt{F_{X}} \simeq M \simeq \Lambda \simeq {\cal{O}}({\rm TeV})$, however, 
it seems particularly difficult
to reliably calculate the SSB parameters.  Furthermore, if there are
no tree-level scalar squared masses, then they may arise from such loops
and be given, roughly, by $M^{2}/16\pi^{2}$ (avoiding a potential
need to introduce a small coefficient $c_{m}$ in front of the squared mass
operators in Table~\ref{table:o1}). 
%

We conclude that, in general, chiral Yukawa couplings
appear once supersymmetry is broken, and if it is broken at low energy
$\sqrt{F_{X}} \simeq M \simeq \Lambda \simeq {\cal{O}}({\rm TeV})$ then
these couplings could be sizable  $y \simeq {\cal{O}}(1)$ yet harmless.
%
%


%
\section{Tree-level Yukawa couplings from the Kahler potential}
\label{sec:tree}
%

In the previous section we classified all supersymmetry
breaking operators and set the supersymmetry breaking scale
parameters to $\sqrt{F_{X}} \simeq M \simeq {\cal{O}}({\rm TeV})$.
Large tree-level Yukawa (and trilinear mass parameters) appear
in that case in the effective theory. Though their parent operators
as well as their order in $F_{X}/M^{2}$ may be different, they are all
{\it a priori} of similar magnitude. 
The issue at hand is therefore not finding possible operators.
Rather, one must avoid excessive mixing between quarks (leptons)
and their mirrors, which could lead to disastrous
contributions to flavor changing neutral currents.
For example, one obvious path one could take 
is to allow tree-level Yukawa couplings
of the same origin (i.e., which are derived from the same operator class)
for all matter fields. This, however, could exactly lead to
such mixing, and furthermore, does
not offer any new insight into the ordinary-mirror fermion mass hierarchy.
We therefore pursue
a more motivated path in which the two sectors
are distinguished by the global symmetries of the effective theory,
and the symmetries induce selection rules which allow/forbid
certain types of Yukawa and  soft 
operators in the different sectors.\footnote{The heaviness of the 
ordinary third family fermions may seem to challenge 
some of the resulting frameworks.
We postpone this discussion to the next section.}
%

This can be done by either exploiting
the global $R$ symmetries which parameterize the hidden supersymmetry
or by symmetries which do not commute with the former symmetries
and therefore characterize the supersymmetry breaking mechanism.
One could also take a linear combination of these choices,
both of which correspond to anomalous symmetries.
In addition, a specific choice of a symmetry is better motivated
if it can provide a hint as for the origin of the ordinary-mirror
fermion mass hierarchy. The model and our parameterizations
already direct one  toward the possible paths:
\begin{itemize}
\item
Recalling that the hard chiral symmetry breaking operators are already 
distinguished by the presence of covariant superspace derivatives, 
which transform under any continuous or discrete $R$-symmetry,
suggests choosing an $R$-symmetry (though this choice is not unique). 
\item
While the $SU(2)_{R}$ symmetry must be broken
(or the fermion and mirror fermion remain degenerate in mass),
the $U(1)_{R}$ of $N=2$ may be preserved and provide the 
desired selection rules. In fact, a $U(1)^{2}$ subgroup of the 
complete $U(2)_{R}$
can survive, where the other $U(1)$ is $U(1)_{J}$.
\item
Mirror parity is a useful tool which enables one to distinguish
matter from mirror matter, and may provide an alternate set of selection rules.
\end{itemize}
%

In order to illustrate the richness of the possible frameworks we
use two distinct sets of selection rules, corresponding to the
symmetry classes mentioned above:
The first group of symmetries is based
on the $N = 2$ preserving
$(A) \, U(1)_{R}^{N = 2} \times Z_{2}^{MP}$;
the second one is based on an Abelian $R$-symmetry
extension of mirror parity $(B) \, U(1)_{R}^{MP}$ which 
explicitly breaks $N = 2$. 
The latter example could be an ``accidental'' symmetry related to
the supersymmetry breaking mechanism. 
We note that it can be mapped to a discrete
$Z_{3} \times Z_{2}$ $R$-symmetry where the $Z_{2}$ is the usual mirror parity
and the chiral coordinate $\theta$ 
and the mirror matter fields all transform as $(1/3)^{-}$.
(Note that once  the transformation properties of one 
matter field and its mirror are fixed,
the $N=2$ superpotential fixes the charge of $\Phi_{V}$, and as a result,
of all other ordinary-mirror bilinears.)  
The symmetry assignments and the corresponding selection rules appear
in tables~\ref{table:sym1} and \ref{table:sym2}.
For illustration, the quark (super)fields $u_{L}$ and $u_{R}$ ($Q$ and $U$) 
and their mirrors are substituted in the operators.
However, we assume identical transformation properties for all quark
and lepton fields so that any other (gauge invariant) combination
of fields could be substituted instead. (It is possible to choose slightly
more complicated examples with (SM-)charge and flavor dependent 
symmetry assignments.) Finally, for completeness we list 
both operators which are holomorphic
(Table~\ref{table:symo}) or non-holomorphic (Table~\ref{table:symo2})
in the Higgs fields, though the latter do not add any intrinsically
new possibilities. Note that it is assumed that only the ordinary Higgs 
doublets, but not their mirrors or any other fields, participate in 
electroweak symmetry breaking. (In particular, mirror parity or its 
extensions are not broken spontaneously by electroweak Higgs VEVs.)
%

A clear tree of possibilities emerges:
\begin{enumerate}
\item
Assume that tree-level mirror-fermion masses arising
from the hard supersymmetry breaking operators,
which occurs naturally in the examples given here.
\item
The chiral symmetries of the ordinary matter fields
may then be broken in the scalar potential, leading
to radiative (ordinary) fermion masses. Alternatively, an effective
$N=1$ Yukawa tree-level superpotential is generated for the ordinary
fields. 
\item
The symmetry properties of both SSB and supersymmetry conserving
operators imply that either both possibilities
for the ordinary fermion mass generation
are allowed or that both are forbidden, as long as the spurion
is not charged under the global $R$-symmetries.
\begin{enumerate}
\item
If both are allowed, a charge assignment for a spurion field
could forbid the supersymmetry conserving operators
and as a result, forbid tree-level masses for the ordinary fermions.
This provides a simple explanation of the matter-mirror
mass hierarchy as a loop factor.  
\item
If both are forbidden, a charge assignment for a spurion field
could allow the supersymmetry conserving operators
and as a result, for tree-level masses for the ordinary fermions.
The ordinary-mirror mass hierarchy can now be explained
by the hierarchy between the charged and neutral spurion
supersymmetry breaking VEVs $F_{X_{1}}/F_{X_{2}}$.
(Note that $\langle F_{X_{2}} \rangle $ itself breaks the $R$-symmetry
if $X_{2}$ is neutral, while $\langle F_{X_{1}} \rangle $ 
may or may not break it.) Alternatively, it could always be that
one class of operators (the hard operators, in this case)
appears at tree level while the other class (the superpotential operators)
appears only radiatively so that the hierarchy is imprinted in the
coefficients of the different operators in $K$.
\end{enumerate}
\end{enumerate}
Many other examples can be constructed along these lines.
%

The symmetry principles nicely arrange the different fermion mass operators.
They also carry implications to most of the other operators.
The scalar squared masses are generically  insensitive
and may arise from tree-level operator with relatively small coefficients 
$c_{m}$, from quadratically divergent loop corrections, or from gauge(ino) 
renormalization. On the other hand
(and similarly to $N=1$ supergravity) 
gaugino mass terms break any Abelian $R$-symmetry,
so that there must be a spurion combination such that
$R(X_{1}X_{2}^{\dagger}) = +2$, consistent with our proposals above.
Our speculation that the $\langle F \rangle$ of the charged spurion
corresponds to a lower scale could lead to suppression of gaugino masses.
Another  group of operator of phenomenological relevance is the operators
corresponding to Higgs mixing at the electroweak scale,
$W \sim \mu\hu\hd$ and $V_{SSB} \sim b\hu\hd + \tilde{\mu}\hinou\hinod$.
Assigning $R(X) = R(\hu\hd)$ always allows for the superpotential
$\mu$-term. (See Table~\ref{table:o1b}.) 
If there is only one spurion, The SSB Higgs (Table~\ref{table:o1}) 
and Higgsino (Table~\ref{table:o2}) mixing operators are independent 
of the (single) spurion charge and cannot be allowed simultaneously.
In the case of a multi-spurion scenario, 
if the  spurions carry different $R$-charges 
then both could co-exist.
(Phenomenologically, both Higgsino mass and 
Higgs mixing in the scalar potential are required in order to
avoid very light Higgs/ino particles in the spectrum.)
%

\begin{table}
\caption{
The $U(1)_{R}^{N = 2} \times Z_{2}^{MP}$ assignment for
the various $N = 1$ superfields. $R(\theta_{\alpha}) = -1$
and all matter superfields are charged as the quark doublet $Q$.
}
\label{table:sym1}
\begin{tabular}{ccccc}
Field & Assignment: && Assignment for mirror: &\\ 
&\\
&Case I&Case II&Case I&Case II\\ \hline
&\\
$H_{i}$ & $0^{+}$ & $-1^{+}$&$0^{-}$ &$+1^{-}$\\
&\\
$Q$ & $+1^{+}$ &$+\frac{1}{2}^{+}$ &$-1^{-}$& $-\frac{1}{2}^{-}$\\
&\\
$\Phi_{V}$ &&& $-2^{-}$ &$-2^{-}$ \\

\end{tabular}
\end{table}
%

\begin{table}
\caption{
The $U(1)_{R}^{MP}$ assignment for
the various $N = 1$ superfields. $R(\theta_{\alpha}) = -1$
and all matter superfields are charged as the quark doublet $Q$.
}
\label{table:sym2}
\begin{tabular}{ccc}
Field & Assignment & Assignment for mirror \\ \hline
&\\
$H_{i}$ & $0$ & $-1$\\
&\\
$Q$ & $0$ & $-1$\\
&\\
&\\ 
$\Phi_{V}$ && $-1$  \\

\end{tabular}
\end{table}
%

\begin{table}
\caption{
Low-energy chiral operators, which are holomorphic in the low-energy fields,
and their symmetry properties. The first, second, and third class of 
operators are soft supersymmetry breaking, hard supersymmetry breaking,
and supersymmetry conserving, respectively.
Allowed operators (assuming $R_{X} = 0$ or $R_{X} = 2$) are underlined. 
}
\label{table:symo}
\begin{tabular}{ccccc}
Operator Class & Operator & Case A1 & Case A2 & Case B\\ 
\hline
&\\
$XX^{\dagger}\Phi^{3}$ & $A\hu\sQ\sU$ & $+2^{+}$& $\underline{0^{+}}$ & $
\underline{0}$ \\  
&\\
& $A\hd\mrsQ\mrsU$ & $-2^{+}$& $-2^{+}$ & $-2$ \\ &\\
& $A\mrhd\sQ\sU$ & $+2^{-}$& $+2^{-}$ & $-1$ \\  &\\
& $A\mrhu\mrsQ\mrsU$ & $-2^{-}$& $0^{-}$ & $-3$ \\ 
\hline
&\\
$XX^{\dagger}\Phi D^{\alpha}\Phi D_{\alpha}\Phi$
&$y\hu u_{L}u_{R}$ & $+4^{+}$& $+2^{+}$ & $+2$\\&\\
&$y\hd\mruL\mruR$ & 
$\underline{0^{+}}$& $\underline{0^{+}}$& $\underline{0}$\\&\\
&$y\mrhd u_{L}u_{R}$ & $+4^{-}$& $+4^{-}$ & $+1$\\&\\
&$y\mrhu\mruL\mruR$ & $0^{-}$& $+2^{-}$ & $-1$\\
\hline
&\\
$X^{\dagger}\Phi^{3}$
&$y\hu u_{L}u_{R}$ & 
$\underline{(+2-R_{X})^{+}}$&$(\underline{0}-R_{X})^{+}$&
$\underline{0}-R_{X}$\\&\\
&$y\hd\mruL\mruR$ & 
$(-2 -R_{X})^{+}$& $(-2 - R_{X})^{+}$& $-2 -R_{X}$\\&\\
&$y\mrhd u_{L}u_{R}$ & $(+2-R_{X})^{-}$& $(+2 -R_{X})^{-}$ & $-1-R_{X}$\\&\\
&$y\mrhu\mruL\mruR$ & $( -2 -R_{X})^{-}$& $(0-R_{X})^{-}$ & $-3-R_{X}$\\

\end{tabular}
\end{table}
%

\begin{table}
\caption{As in Table \ref{table:symo} but for
operators which contain {$\Phi^{\dagger}$}.
}
\label{table:symo2}
\begin{tabular}{ccccc}
Operator Class & Operator & Case A1 & Case A2 & Case B\\ 
\hline
&\\
$XX^{\dagger}\Phi^{\dagger}\Phi^{2}$& 
${\cal{A}}\hd^{\dagger}\sQ\sU$ & $+2^{+}$& $+2^{+}$ & $\underline{0}$\\&\\
& ${\cal{A}}
\hu^{\dagger}\mrsQ\mrsU$ & $-2^{+}$& $\underline{0^{+}}$ & $-2$ \\ &\\
& ${\cal{A}}\mrhu^{\dagger}\sQ\sU$ & $+2^{-}$& $0^{-}$ & $+1$ \\  &\\
& ${\cal{A}}\mrhd^{\dagger}\mrsQ\mrsU$ & $-2^{-}$& $-2^{-}$ & $-1$ \\ 
\hline
&\\
$XX^{\dagger}\Phi^{\dagger} D^{\alpha}\Phi D_{\alpha}\Phi$
&$y\hd^{\dagger} u_{L}u_{R}$ & $+4^{+}$& $+4^{+}$ & $+2$\\&\\
&$y\hu^{\dagger}\mruL\mruR$ & 
$\underline{0^{+}}$& $2^{+}$& $\underline{0}$\\&\\
&$y\mrhu^{\dagger} u_{L}u_{R}$ & $+2^{-}$& $+2^{-}$ & $+3$\\&\\
&$y\mrhd^{\dagger}\mruL\mruR$ & $0^{-}$& $0^{-}$ & $+1$\\
\hline
&\\
$X^{\dagger}\Phi^{\dagger}\Phi^{2}$
&$y\hd^{\dagger} u_{L}u_{R}$ & 
$\underline{(+2-R_{X})^{+}}$&$\underline{(+2-R_{X})^{+}}$&
$\underline{0}-R_{X}$\\&\\
&$y\hu^{\dagger}\mruL\mruR$ & 
$(-2 -R_{X})^{+}$& $(\underline{0} - R_{X})^{+}$& $-2 -R_{X}$\\&\\
&$y\mrhu^{\dagger} u_{L}u_{R}$ & 
$(+2-R_{X})^{-}$& $ (0-R_{X})^{-}$ & $+1-R_{X}$\\&\\
&$y\mrhd^{\dagger}\mruL\mruR$ & 
$( -2 -R_{X})^{-}$& $(-2-R_{X})^{-}$ & $-1-R_{X}$\\

\end{tabular}
\end{table}
%
%


%
\section{A heavy generation}
\label{sec:mixing}
%

In our discussion so far we distinguished ordinary from mirror matter,
but did not distinguish, for example, light and heavy SM (ordinary) fermions.
That is, if one of the mechanisms to render ordinary fermions light relative
to the mirror fermion is realized, then {\it all} 
of the ordinary fermions will be light with masses of
roughly the same order of magnitude. 
However, the SM fermion spectrum contains two special cases:
The first case is that of the top quark 
(or for that matter, of all of the third family)
whose mass is of the order of the mirror fermion masses. 
The second case is that of the nearly massless neutrinos. 
We postpone the discussion
of the neutrinos to the next section and focus here on the case
of heavy SM fermions. 
%

While in some cases internal hierarchy within
the SM sector can be put in by hand, it is not always sufficient.
For example, if the SM fermion mass is generated radiatively,
vacuum stability constraints make it very unlikely that the top ($\tau$)
in the quark (lepton) sector receives its mass radiatively (with a large
trilinear parameter put in by hand).
This would require hard quartic couplings of order $\kappa \gtrsim 4\pi$.
An alternative tree-level mechanism may exist,
particularly in the latter case.
One obvious candidate for such a mechanism
is mirror-symmetry breaking in the third family
and consequently, mass mixing between ordinary and mirror third 
family fermions. As long as such mixing is constrained to only the 
third family, the implications to flavor changing neutral currents are 
generically within experimental constraints. 
Mirror parity breaking in such a scenario is intimately linked
to the flavor symmetry structure. We first discuss the phenomenology
of such a mechanism, and then speculate on its possible origin
from a spontaneously broken Abelian flavor (gauge) symmetry. 
%

If one allows MPV in the third family, 
then there could be tree-level mixing between the fermions and their mirrors.
For explicitness, let us concentrate on the case of 
the top quark and its mirror with mixing terms: $\tilde{\mu}_{L}'t_L\mrtL$ and
$\tilde{\mu}'_Rt_R\mrtR$. For simplicity, let us further assume that
the usual quark mass term $t_Lt_R$ is small and can be taken to be zero.  
The mirror top quark, on the other hand, 
has a mass term $M\mrtL\mrtR$, which is assumed to arise at tree level 
and $M \simeq \mweak$.  ($M$ here is not the supersymmetry breaking scale
but simply the large mass parameter in the fermion mass matrix,
$M \equiv M_{f_{L}^{\prime}f_{R}^{\prime}}$.)
A similar structure holds for the bottom sector, with identical 
$\tilde{\mu}_{L}'$ (from the SM $SU(2)_{L}$ symmetry) 
for the left-handed bottoms
but with independent  $\tilde{\mu}'_R$ and $M$ parameters.
%

Defining 
\begin{equation}
\psi^+_j=\left(
\begin{array}{c}t_L\\\mrt_R\end{array}
\right), \ \ \ 
\psi^-_j=\left(
\begin{array}{c}\mrt_L\\t_R\end{array}
\right), \ \ \ j=1,2, 
\end{equation}
the mass matrix can be written as 
\begin{equation}
(\psi^+\ \psi^-)\left(
\begin{array}{cc}
0&X^{\rm T}\\X&0\end{array}\right)
\left(\begin{array}{c}\psi^+\\\psi^-\end{array}
\right)+{\rm H.c.}, 
\end{equation}
where
\begin{equation}
X=\left(
\begin{array}{cc}
\tilde{\mu}_{L}'&M\\0&\tilde{\mu}^{\prime}_R\end{array}\right),
\label{Xmatrix}
\end{equation}
and we neglected a pure SM top mass. (The MPV mixing may
be SSB or $N=1$ supersymmetric, though here we use the SSB notation.)
The mass eigenstates $\chi^{\pm}$ are readily found,
\begin{equation}
\chi^+_i=V_{ij}\psi^+_j=\left(
\begin{array}{cc}
\cos\phi_+&\sin\phi_+\\
-\sin\phi_+&\cos\phi_+\end{array}\right)\left(
\begin{array}{c}t_L\\\mrt_R\end{array}
\right),\ \ 
\chi^-_i=U_{ij}\psi^-_j=\left(
\begin{array}{cc}
\cos\phi_-&\sin\phi_-\\
-\sin\phi_-&\cos\phi_-\end{array}\right)\left(
\begin{array}{c}\mrt_L\\t_R\end{array}
\right).
\end{equation}
Here $U$ and $V$ are the unitary matrices chosen to diagonalize the mass 
matrix:
\begin{equation}
U^*XV^+=M_{\rm Dirac}.
\end{equation}
The mass eigenvalues $M_{{\rm Dirac}_{1,2}}^2$ are
\begin{equation}
M_{{\rm Dirac}_{1,2}}^2=\frac{M^2+{\tilde{\mu}}^{\prime{2}}_{L}+
\tilde{\mu}^{\prime{2}}_R\mp
\sqrt{(M^2+{\tilde{\mu}}^{\prime{2}}_{L}
+\tilde{\mu}^{\prime{2}}_R)^2-4{\tilde{\mu}}^{\prime{2}}_{L}
\tilde{\mu}^{\prime{2}}_R}}{2},
\end{equation}
while the mixing angles $\phi^+$ and $\phi^-$ can be written as 
\begin{eqnarray}
\tan\phi_+&=&\frac{\tilde{\mu}^{\prime{2}}_R+M^2-{\tilde{\mu}}^{\prime{2}}_{L}
-\sqrt{(M^2+{\tilde{\mu}}^{\prime{2}}_{L}+\tilde{\mu}^{\prime{2}}_R)^2-4
{\tilde{\mu}}^{\prime{2}}_{L}\tilde{\mu}^{\prime{2}}_R}}
{2{\tilde{\mu}}^{\prime}_{L}M}\\
&& \nonumber \\
\tan\phi_-&=&\frac{\tilde{\mu}^{\prime{2}}_R-M^2-{\tilde{\mu}}^{\prime{2}}_{L}-\sqrt{(M^2+
{\tilde{\mu}}^{\prime{2}}_{L}+\tilde{\mu}^{\prime{2}}_R)^2-4{\tilde{\mu}}^{\prime{2}}_{L}\tilde{\mu}^{\prime{2}}_R}}
{2\tilde{\mu}^{\prime}_RM}.
\end{eqnarray}
%

The mass splitting between the ordinary and mirror quarks 
is a function of  ${\tilde{\mu}}^{\prime}_{L}$, $\tilde{\mu}^{\prime}_R$ 
and $M$. Two limits are of particular interest:
\vspace*{0.2cm}

{\it 
$\bullet \,\,\, \tilde{\mu}^{\prime}_{L}$, $\tilde{\mu}^{\prime}_R$ 
and $M$ are all of the same order of magnitude:
}\vspace*{0.2cm}

Assume, as an example,  ${\tilde{\mu}}^{\prime}_{L}=\tilde{\mu}^{\prime}_R=M$.
In this limit, $M_{{\rm Dirac}_{1,2}}$ are of the same order 
of magnitude and there is large mixing between the ordinary SM quarks 
and their mirror partners: 
\begin{equation}
M_{{\rm Dirac}_{1,2}}
=\left(\frac{3\mp\sqrt{5}}{2}\right)^{\case{1}{2}}M=\left\{
\begin{array}{c}0.62\ M\\1.62\ M\end{array}\right.,
\ \ \tan\phi_{\pm}=\frac{-\sqrt{5}\pm{1}}{2}.
\end{equation}
This case is relevant for the top sector.  The top quark can
get its large mass while the mirror top is sufficiently heavy to evade current
experimental limits that may apply. 
\vspace*{0.2cm}

{\it 
$\bullet$ \,\,\, The MPV mixing between the ordinary 
quarks and the mirror partners is much smaller than $M$:
}\vspace*{0.2cm}

Assume, without loss of generality, $\tilde{\mu}^{\prime}_R\ll{M}$. 
One has, to leading order in $\tilde{\mu}^{\prime}_R/M$,
\begin{eqnarray}
M_{{\rm Dirac}_{1}}&=&\frac{\tilde{\mu}^{\prime}_R{\tilde{\mu}}^{\prime}_{L}}{\sqrt{M^2+
{\tilde{\mu}}^{\prime{2}}_{L}}}
\stackrel{{\tilde{\mu}}^{\prime}_{L}\ll{M}}{=}
\frac{{\tilde{\mu}}^{\prime}_{L}\tilde{\mu}^{\prime}_R}{M},\\
M_{{\rm Dirac}_{2}}&=&
\sqrt{M^2+{\tilde{\mu}}^{\prime{2}}_{L}}
\left(1+\frac{\tilde{\mu}^{\prime{2}}_RM^2}{2(M^2+{\tilde{\mu}}^{\prime{2}}_{L})^2}\right)
\stackrel{{\tilde{\mu}}^{\prime}_{L}\ll{M}}{=}M\left(1+
\frac{{\tilde{\mu}}^{\prime{2}}_{L}+\tilde{\mu}^{\prime{2}}_R}{2M^2}\right)
\end{eqnarray}
The mixing angles in this limit can be similarly obtained and read
\begin{eqnarray}
\tan\phi_+&=&-\frac{{\tilde{\mu}}^{\prime}_{L}}{M}(1-
\frac{\tilde{\mu}^{\prime{2}}_R}{M^2+
{\tilde{\mu}}^{\prime{2}}_{L}})
\stackrel{{\tilde{\mu}}^{\prime}_{L}\ll{M}}{=}-\frac{{\tilde{\mu}}^{\prime}_{L}}{M}(1-\frac{\tilde{\mu}^{\prime{2}}_R}{M^2})\\
\tan\phi_-&=&-\frac{(M^2+{\tilde{\mu}}^{\prime{2}}_{L})}
{M\tilde{\mu}^{\prime}_R}(1-\frac{{\tilde{\mu}}^{\prime{2}}_{L}
{\tilde{\mu}}^{\prime{2}}_{R}}
{(M^2+{\tilde{\mu}}^{\prime{2}}_{L})^2})
\stackrel{{\tilde{\mu}}^{\prime}_{L}\ll{M}}{=}-
\frac{M}{{\tilde{\mu}}^{\prime}_{R}}(1+
\frac{{\tilde{\mu}}^{\prime{2}}_{L}}{M^2}).
\end{eqnarray}
As one expects,
one of the eigenstates becomes light when one of the mass mixing is small, 
while the heavy mass eigenvalues is still $\sim{\cal{O}}(M)$.  
The fraction of the usual right-handed quark  in the light eigenstate, 
$\sin\phi_-$, is always large since $\tan\phi_-$ is much 
larger than 1.  However, the fraction of the usual left-handed 
quark, $\cos\phi_+$, depends on the ratio of ${\tilde{\mu}}^{\prime}_{L}/M$.  
It is large when ${\tilde{\mu}}^{\prime}_{L}$ is much smaller than $M$. 
Alternatively, one can have large mixing between
the ordinary and mirror quarks when ${\tilde{\mu}}^{\prime}_{L}$ and $M$ 
are of the same order. 
A similar situation happens when 
${\tilde{\mu}}^{\prime}_R$ is of the same order of magnitude as $M$ while 
${\tilde{\mu}}^{\prime}_{L}$ is much smaller. 
%

The latter limit enables one to realize simultaneously a heavy  
top quark mass and a few GeV bottom quark mass.
The parameter ${\tilde{\mu}}^{\prime}_{L}$ is the same for both
top and bottom sector and should be of the order of $M_{\mrt_L\mrt_R}$ 
so that the top is sufficiently heavy.
However, ${\tilde{\mu}}^{\prime}_R$ and $M$ could be different for the two
sectors.  As long as $({\tilde{\mu}}^{\prime}_R/M)_{b}$ is ``small'', 
the contribution to the  bottom quark mass is ``small''.   
The most attractive choice is to have 
$M_{\mrb_L\mrb_R}\sim{M}_{\mrt_L\mrt_R}\sim{{\tilde{\mu}}^{\prime}_{L}}\gg({\tilde{\mu}}^{\prime}_R)_b$. Another possibility is to take 
$M_{\mrb_L\mrb_R}\gg{M}_{\mrt_L\mrt_R}\sim{{\tilde{\mu}}^{\prime}_{L}}
\sim({\tilde{\mu}}^{\prime}_R)_b$. 
This is, however, more difficult to realize since
it is difficult to obtain such a large value for $M_{\mrb_L\mrb_R}$
which is proportional to the Higgs VEV.
%

We conclude that once MPV mixing is allowed, it is possible to realize
heavy and highly mixed ordinary and mirror top quarks
simultaneously with a relatively light (and relatively non-mixed)
SM bottom quark. The question we would like to consider next is
with regard to the possible mechanisms that give rise to such a mixing. 
Various possibilities exist, for example, a ``flavored spurion'' such that
$X^{\dagger}Q_{i}\mrQ_{j}$ terms are allowed in the Kahler potential
for $i=j=3$. Mirror symmetry could be viewed in this case an accidental
symmetry of the first two generations or as a flavor symmetry.
(Note that only vector-like mixing terms are allowed by the SM
gauge symmetries.)
Here, however, we will present a different toy model
in which mirror symmetry breaking is a result
of a spontaneous breaking of a gauged flavor symmetry.
%

Assume an additional (horizontal) 
$U(1)_H$ gauge factor. The superpotential contains, 
for example, the  term  $g_{H}h_{Q}Q_3\Phi_H\mrQthird$, 
assuming that $Q_3$ and $\mrQthird$
are charged under the horizontal gauge symmetry with charge $\pm h_{Q}$,
and $g_{H}$ is the horizontal gauge coupling.  
$\Phi_H$ is a gauge singlet
contained in the $U(1)_H$ $N =2$ vector multiplet.  
If it develops a  VEV $\langle\Phi_H\rangle$, it would create a mixing 
parameter  $\mu_{Q}^{\prime} = g_{H}h_{Q}\langle\Phi_H\rangle$.
In this example, the Kahler potential can
still preserve the mirror symmetry, which is broken spontaneously 
by  $\langle\Phi_H\rangle$. 
(Note that the flavor symmetry itself is not broken by the 
$\langle\Phi_H\rangle$ VEV.)
This proposal provides a simple framework for the generation
of the mixing terms employing generation-dependent $U(1)$ symmetries.
However, one must overcome certain difficulties before 
such a proposal can be realized. We outline those difficulties and
the possible cures below.
%

First, the relative size of the mixing parameter is proportional
to the hypermultiplet horizontal charge.
It may not be straightforward to find an anomaly-free combination
that naturally produces the desired hierarchy.
Certain fields, however, could be singlets (for example, $D_{3}$).
Also, a combination of different $M$ parameters could also contribute
to the hierarchy.  
%

Secondly, there is the issue of the mixing of the
third generation quarks with
the two light generation quarks. If the third family quarks are charged
under any symmetry while the light quarks and the Higgs bosons are neutral,
then any inter-family mixing is forbidden. This can be resolved, for example,
by the introduction of a SM singlet hypermultiplet $S$ which is also charged
under $U(1)_H$ so that an appropriate chiral symmetry breaking 
term is allowed in Kahler potential, e.g., $X^{\dagger}S\hd Q_{3}D_{2}$,
which could lead to an $A$-term or a Yukawa coupling proportional
to $\langle F_{S} \rangle$ and $\langle S \rangle$, respectively.
In the case that only $A$-terms arise then the intergenerational
mixing is naturally suppressed as the square of the loop factor.
The $S$-VEVs are induced by the dynamics below the supersymmetry
breaking scale, e.g., the SSB potential, 
and are therefore suppressed. A $S$-VEV breaks
the horizontal symmetry spontaneously and its size is constrained
by the usual considerations related to the presence
of an extra neutral $Z^{\prime}$ gauge boson \cite{moreZ}. 
%

Lastly, consider the operators $X^{\dagger}\Phi_{H}Q_{2}\mrQ_{2}$,
$XX^{\dagger}\Phi_{H}D^{\alpha}Q_{2}D_{\alpha}\mrQ_{2}$ etc. While the 
former operator can be forbidden by the $R$-symmetry, the latter is
allowed by the symmetries.
If such operators arise they could lead to ordinary-mirror matter 
mixing in all three generations. 
The supersymmetry dynamics must therefore be constrained not
to generate such vector-hypermultiplet mixed operators.
%

The  proposed toy model serves to illustrate that the flavor 
symmetry may be intimately linked to the details of the breaking 
of supersymmetry and of the global symmetries it induces.  
In particular, the heaviness of the third family may stem
from the heaviness of the mirror fermions, in which case
either mirror symmetry plays the role of a flavor symmetry
or the flavor symmetry breaks the mirror symmetry.
%
%


%
\section{Neutrino masses}
\label{sec:neutrino}
%

Recent results from the atmospheric and solar neutrino oscillation 
experiments indicates non-zero neutrino masses, although extremely 
small with respect to the charged leptons \cite{neutrino}.
The mass squared difference between two neutrino mass eigenstates is of the 
order of 
${10}^{-3}\ {\rm eV}^2$ from atmospheric neutrino oscillation data (and 
${10}^{-5}\ {\rm eV}^2$ for solar neutrinos).  
The smallness of neutrino masses can be explained most simply by the see-saw
mechanism\cite{seesaw}, where a right-handed sterile 
neutrino $N_{R}$ with a Majorana mass $M$ is introduced.
(Again, $M$ here is the large mass parameter in the fermion mass matrix,
$M \equiv M_{N}$.) Assuming a Dirac mass $m_D N_{R}\nu_L$, 
and that there is no Majorana mass for the left-handed neutrinos, 
the light mass eigenvalue is $m\sim{m}_D^2/M$.  For $m_D$ of the 
order of the electroweak scale, the tiny neutrino masses can be 
obtained if $M \sim 10^{15}$ GeV is of the order of the unification scale.  
%

The extended neutrino sector in the MN2SSM is 
strongly constrained by experiment.
There are six active neutrinos, the three ordinary neutrinos $\nu$ 
and their mirror partners $\nu^{\prime}$,  
all of which couple to the electroweak gauge bosons.
Given the constraints from the invisible $Z$-boson width on the number
of active neutrinos in $Z$ decays, $N_\nu=2.994\pm{0.012}$ \cite{pdg},
the mirror neutrinos cannot be light:
Any additional active neutrinos such as $\nu^{\prime}$
must be heavier than $m_Z/2$.  
(If there are additional sterile neutrinos, there could be more than three 
light active neutrinos as the coupling $Z\bar{\nu}_i\nu_i$, 
$\nu_i$ being the mass eigenstate, 
is suppressed by mixing angles.  Nevertheless, we assume only three 
light active neutrinos.)
As an obvious consequence from the last statement, the observed neutrino 
oscillations cannot be explained by  $\nu\rightarrow{\nu}^{\prime}$ and must
occur among the ordinary neutrinos. 
In the following, we will only
address the question of obtaining the small ordinary neutrino masses
while keeping the mirror neutrinos heavy.  
The mixing between the light neutrino mass eigenstates and an explanation 
of the oscillation data require a more careful model building 
(for example, the generational structure of the matrices discussed
below needs to be addressed), which is left to future studies.
We therefore discuss only one generation of neutrinos.
%

Clearly, the see-saw mechanism described above does not generalize
to  $N=2$ supersymmetry:
If the small neutrino mass is generated by the usual seesaw
mechanism, the sterile neutrino must be heavy, with its mass above  
the $N=2$ breaking scale. The
mirror sterile  neutrino  must have the same mass 
because the exchange symmetry is a good symmetry
above the $N=2$ breaking scale.    
The mirror neutrino masses are then also
suppressed by the see-saw mechanism,
$m_{\nu^{\prime}} \sim \langle H \rangle^{2}/M \ll m_{Z}/2$,
which is below experimental bounds. Therefore, the Majorana
mass for the  sterile neutrino cannot be much larger than the $N=2$
breaking scale, which in the framework of $2\times 2$ see-saw
leads to heavy neutrinos (unless the Yukawa couplings are fine-tuned).
(We note in passing, that if the three right-handed neutrinos remain
as light as the the left-handed neutrinos, one can explore
an explanation of the oscillation data involving also
$\nu_{L_{i}} \rightarrow N_{R_{j}}$ transitions.)
%

If mirror parity is conserved, there is no mixing between the usual 
and mirror neutrinos.  The mass matrix is reduced to two diagonal blocks
for the usual and mirror sectors.  Once sterile neutrinos are
introduced the mirror neutrino mass can be
generated via  effective tree level Yukawa coupling as for the other
mirror fermions. The ordinary neutrino mass, on the other hand, 
cannot be generated  radiatively since the right-handed
neutrino is a gauge singlet.  
Common techniques like the radiative generation 
of neutrino masses via the introduction of a charged $SU(2)$-singlet
and a second Higgs doublet, or tree-level neutrino mass by a 
Higgs triplet\cite{neutrinomass}
may  be exploited to give the small neutrino masses, 
particularly since such fields are available in the spectrum.
(One could also introduce by hand 
tiny tree-level effective Yukawa couplings for the neutrinos
or extend the SM group as discussed in Section \ref{sec:loop}.)
Here, we will explore a different source of
neutrino masses in $N=2$ scenarios, a $3\times 3$ see-saw mechanism
which is induced by a small breaking of the mirror parity.
%

With only the minimal spectrum (no sterile neutrinos)
but with MPV mixing between the usual and mirror neutrinos 
$\tilde{\mu}^{\prime}_{\nu}\nu\nu'$,
the neutrino mass matrix is given by
\begin{equation}
\left(
\begin{array}{cc}
0&\tilde{\mu}^{\prime}_{\nu}\\
\tilde{\mu}^{\prime}_{\nu}&0
\end{array}
\right)
\end{equation}
in the basis of $(\nu,\nu')$.  
In this simplest framework one has
two degenerate mass eigenstates 
and no mass hierarchy between the mass eigenstates can be generated.  
%

Let us then consider a more general $N=2$ neutrino sector.
(For simplicity, we only consider one generation.) Consider
\begin{itemize}
\item
Two sterile neutrino superfields, $N$ and its mirror partner $N'$,
with masses $M_N$ and $M_N'$, respectively. (We omit hereafter the $R$ index.)
\item
Dirac masses for the usual and mirror sector $m_{D}{N}\nu$, 
$m_D'{N}'\nu'$.
\item
Mirror parity violating terms 
$\tilde\mu^{\prime}_{\nu}\nu\nu'$, $\tilde{\mu}^{\prime}_NNN'$.
\item
Dirac-type mixing $\tilde\mu^{\prime}_{N\nu'}N\nu'$, 
$\tilde{\mu}^{\prime}_{N'\nu}N'\nu$.
\end{itemize}
Under these assumptions the $4\times4$ neutrino mass matrix reads
\begin{equation}
(\nu,N,\nu',N')\left(
\begin{array}{cccc}
0&m_D&\tilde\mu^{\prime}_{\nu}&\tilde\mu^{\prime}_{N'\nu}\\
m_D&M_N&\tilde\mu^{\prime}_{N\nu'}&\tilde\mu^{\prime}_{N}\\
\tilde\mu^{\prime}_{\nu}&\tilde\mu^{\prime}_{N\nu'}&0&m_D'\\
\tilde\mu^{\prime}_{N'\nu}&\tilde\mu^{\prime}_{N}&m_{D}'&M_{N}'
\end{array}
\right)
\left(
\begin{array}{c}
\nu\\N\\\nu'\\N'
\end{array}
\right).
\end{equation}
%

However, it is simplified under a well-motivated set of assumptions
which we consider for the purpose of illustration: 
\begin{itemize}
\item
There is no Dirac mass in the usual neutrino sector, $m_D=0$.  
This is true, for example, if the ordinary sector 
fermion masses originate from radiative corrections.
\item
There is no mixing between the sterile neutrinos $N$ and $N'$, 
$\tilde\mu'_N=0$.  
This is the case if the mixing arises from the VEV of some 
mirror $U(1)$ gauge boson singlet 
$\langle\Phi_{U(1)_{\nu}}\rangle$, while $N$ and $N'$ are singlets 
under $U(1)_{\nu}$. 
\item
There is no Dirac-type mixing,  $\tilde\mu^{\prime}_{N\nu'} =
\tilde{\mu}^{\prime}_{N'\nu}=0$.  
Assuming that Higgs couplings preserve mirror parity,
those terms could only arise from
Yukawa terms in the Kahler potential involving the superfield
combinations $H_1'LN'$, $H_2^{\prime{\dagger}}LN'$  
and $H_2'L'N$, $H_1^{\prime{\dagger}}L'N$, once the mirror Higgs bosons 
acquire VEVs. However, we assume that EWSB is induced only by
the ordinary MSSM Higgs doublets, so 
$\langle{H}_1'\rangle = \langle{H}_2'\rangle=0$ 
and such Yukawa terms do not generate mixing.
\end{itemize}
The usual sterile neutrino now decouples,
and one is  left with a $3\times 3$  mass matrix with only 
three parameters:
\begin{equation}
\left(
\begin{array}{ccc}
0&\tilde\mu^{\prime}_{\nu}&0\\
\tilde\mu^{\prime}_{\nu}&0&m_D'\\
0&m_{D}'&M_N'
\end{array}
\right).
\end{equation}
%

In the relevant limit one has $\tilde\mu'_{\nu}\ll{m}_D'\ll{M}_N'$, 
i.e., the MPV parameter $\tilde{\mu}'_\nu$ is small and 
the Dirac mass for the mirror neutrinos,
$m_D'\sim\langle{H}\rangle$, is small with respect to 
sterile neutrino mass $M_N'$
which is of the order of the $N=2$ breaking scale.
The mass eigenvalues in this limit are approximately
\begin{equation} 
m_1\sim\frac{{\tilde\mu}^{\prime{2}}_{\nu}}{m_2},
\ \ \ m_2\sim\frac{{m_D'}^2}{M_N'},
\ \ \ m_3\sim{M}_N'.
\end{equation}
The smallness of the lightest neutrino masses 
can be controlled by the small MPV parameter
${\tilde\mu}^{\prime}_{\nu}$, 
while the second lightest neutrino remains  heavy as long as 
${m_D}'/{M_N'}$ is not too small.  
It is, however, in the mass range implied by electroweak data
(see the next section) and a candidate for the LMP.
Notice that $M_N'$ cannot be 
too large or the mirror neutrino mass would be suppressed
below the experimental  lower limit. As an example, taking 
$M_N'=1000$ GeV, ${m_D}'=300$ GeV and 
${\tilde\mu}^{\prime}_{\nu}
\sim{10}^{-6}-10^{-4}$ GeV, the neutrino masses read:
\begin{equation}
m_1\sim{10}^{-5}-10^{-1}\ {\rm eV},\ \ m_2\sim{90}\ {\rm GeV},\ \ 
m_3\sim{1000}\ {\rm GeV}.
\end{equation}
This model is a variation of  the see-saw mechanism where the small 
mirror parity violating parameter ${\tilde\mu}^{\prime}_{\nu}$ 
plays the role of the usual Dirac masses.  
%

In conclusion,
The neutrino sector in $N=2$ supersymmetry 
is strongly constrained as one needs to not only  
generate the small neutrino masses to fit the neutrino 
oscillation data, but also to maintain  
a large mass hierarchy between the ordinary and mirror sectors. 
Here, we presented a simple model for the  $N=2$ neutrino sector
which relies on small MPV. The  model is successful
though it is far from unique
and other possibilities need to be explored.
%
%


%
\section{Phenomenology of $N=2$ Supersymmetry}
\label{sec:pheno}
%

Given its extended spectrum,
the phenomenology of the MN2SSM is particularly rich.
Its effects are both indirect (electroweak physics)  
and direct (collider phenomenology and new particle searches).
Although many predictions depend on the details of the model,
important conclusions can be drawn based only 
on the general structure of the $N=2$ framework.
While some of the MN2SSM characteristics only provide  
a variation on the  phenomenology of the
$N=1$ MSSM, many other features are unique to $N=2$, 
and provide the smoking gun signals for 
the discovery of $N=2$ supersymmetry. 
In this section
we review some of the more interesting aspects, both indirect and direct,
of $N=2$ supersymmetry.
\subsection{Electroweak and Higgs Physics}
%

As mentioned in the introduction, additional chiral quarks and leptons 
can lead to  large positive contribution to the 
oblique parameter $S$ \cite{S}, 
which is phenomenologically unfavorable, if not excluded \cite{pdg}.  
However, the well known result that each new fermion 
generation leads to $+2/3\pi$ contribution to $S$ \cite{S} holds only if
the extra generation is degenerate in mass with $m_{f_{\rm new}}\gg{m}_Z$.  
In the case of $N=2$ and the MN2SSM, 
the masses of the mirror matter fermions are related to the EWSB
Higgs VEVs, and so $m_{f^{\prime}} \simeq m_Z$. 
Furthermore, the origin of its mass is similar to that of the 
ordinary fermions (supersymmetry breaking operators in
the Kahler potential) and there is no reason to assume degeneracy.
As shown in Ref.~\cite{oblique1}, electroweak precision data can
accommodate extra generations  
if there exist heavy (active) neutrinos with masses close to 50 GeV
(while their charged $SU(2)_{L}$ partners are with 
masses slightly above $100$ GeV).
This is because mass dependent terms become important once the fermions
are relatively light and their non-degenerate
spectrum breaks the custodial $SU(2)$ symmetry of the electroweak interactions.
This scenario can be naturally fulfilled in the MN2SSM:
An example of a mirror neutrino in this mass range was given in
the previous section.
In addition, Ref.~\cite{oblique1} has also found that 
extra generations may be accommodated
if charginos and neutralinos have masses close to 60 GeV.
This is an example of negative contributions to $S$ from
custodial $SU(2)$ breaking Majorana masses \cite{oblique2}.
In $N=2$ such Majorana masses arise
naturally, for example, there are mixing terms between
the mirror Higgsino doublets and $\psi_{W}$.
This again can lead to a negative or vanishing value of $S$.\footnote{
Additional gauged (flavor) $U(1)$, as suggested 
in Section~\ref{sec:mixing},
can also contribute negatively to $S$, 
depending on the mixing between the extra gauge boson $Z'$
and the ordinary $Z$ \cite{moreZ}.}
We conclude that while consistent,
the $S$ parameter places probably the strongest  
constraints on the MN2SSM. It requires some relatively light
mirror particles, for example,
some combination of
relatively light mirror neutrinos and mirror Higgsinos.
A dedicated electroweak analysis including all mirror particles
is well motivated.
%

Another issue of importance to electroweak physics is
the mass of the Higgs boson.
The MN2SSM Higgs sector is not as constrained as in the MSSM
or other $N=1$ frameworks. 
The number of Higgs doublets participating in EWSB
could vary, in principle, between one to four.\footnote{ 
Note that the $N=1$ anomaly cancellation 
requirement of an even number of Higgs superfields
is automatically satisfied in $N=2$ for any number of
Higgs hypermultiplets.
The $N=1$ requirement of at least two  Higgs superfields
with opposite hypercharge acquiring VEVs (from fermion mass generation)
is relaxed in the framework of low-energy supersymmetry
breaking and the MN2SSM due to the appearance of non-holomorphic
Yukawa terms. Therefore, there could be, in principle, 
only one Higgs hypermultiplet
with only the Higgs, not its mirror, acquiring a VEV.}
Here we assume two, $\hd$ and $\hu$, as in the MSSM.
Even within this MSSM-like framework of two Higgs doublets
participating in EWSB, there is no upper
bound on the mass of lightest Higgs boson.
This is because
tree-level Higgs quartic couplings $\lambda$ arise not only from supersymmetric
terms  $\lambda \sim g^{2}$ as in $N=1$, but also 
from hard supersymmetry breaking operators in the Kahler potential 
$\lambda \sim g^{2} + \kappa$. (See Section~\ref{sec:hard}.)
Consequently, the minimization
of the Higgs potential leads to a modified
formula for relating $m_Z$, $m_{H_{1,2}}^2$, $\tan\beta$, and 
$\mu$, as in Eq.~(\ref{minimization}). More importantly, 
because of its dependence on the arbitrary hard couplings $\kappa$,
the tree-level light Higgs mass 
$m_{h}^{2} \sim \lambda(\nu_{1}^{2} + \nu_{2}^{2})$ 
is not bound from above by $m_{Z}$ (or by $130$ GeV at loop order)
as in the MSSM.
This observation is not unique to $N=2$ but rather
to theories with low-energy supersymmetry breaking where 
$\kappa \sim {\cal{O}}(1)$ is possible. It carries important implications
for defining theoretically motivated mass range for future Higgs searches.
%

Another indirect implications arise from the fact that
the ordinary quark and lepton masses (except the third 
generation) may arise radiatively, 
which by itself has interesting consequences \cite{BFPT}. 
A general feature in theories with 
radiative fermion masses is that the 
anomalous magnetic moments are not suppressed
by a  loop factor relative to the respective fermion mass: 
$a_f\sim{m_f^2}/{\tilde{m}}^2$, where $\tilde{m}$ is the mass of the heavy 
particles running in the loop.  
This has particular relevance in the case of the muon
whose magnetic moment is well measured and further
improvement in its measurement is expected in the near future \cite{muon},
allowing for such effects to be observed \cite{BFPT}.
%

The light mirror fermions, the possibility
of hard Yukawa and quartic couplings as well as of radiative Yukawa
couplings, and the large number of new degrees of freedom,
all imply that the MN2SSM interacts with the SM and electroweak physics
more strongly than the MSSM and leads to quite different predictions
for various observables.
\subsection{Collider Phenomenology}
%

The experimental limits on the extra heavy quarks and leptons are based on 
searches for the fourth generation at $e^+e^-$ and $p\bar{p}$ 
colliders: $m_{\nu'}\gtrsim {40}$ GeV,  $m_{l'}\gtrsim {80}$ GeV, 
and $m_{b'}\gtrsim 128$ GeV \cite{pdg}.  These lower mass bounds 
may or may not apply  to the $N=2$ mirror quarks and leptons, 
as they depend on the decay modes of the heavy fermions.  
Nevertheless, such limits  are easily satisfied
for $\sqrt{\langle{F_{X}}\rangle}\sim{M}$,
corresponding to effective tree level Yukawa couplings for the mirror fermions 
of the order of unity. On the other hand,
the mirror fermion masses are proportional to the 
EWSB Higgs VEV and therefore cannot be much 
larger than the electroweak scale $m_{f^{\prime}} \simeq
\langle H \rangle = 174$ GeV. 
(In addition, the oblique $S$ parameter 
also constraints some of the masses from above.)
This upper bound ensures that mirror fermions can be 
copiously produced at any machine that produces
a large number of top pairs: The Tevatron, CERN's Large Hadron Collider
(LHC) and future lepton colliders.
The MN2SSM and the $N=2$ framework
can be confirmed or excluded shortly after the next energy 
frontier is reached.
%

The experimental signals largely depend on whether the mirror 
parity $M_{P}$ and/or the usual $R$-parity $R_{P}$ are broken 
below the $N=2$ breaking scale.   
Given both parities, there are three special particles that
play an important role in determining the phenomenology:
the lightest $R_{P}$-odd (supersymmetric) particle (LSP), 
the lightest $M_{P}$-odd (mirror) particle (LMP), 
and the lightest mirror supersymmetric particle (LMSP), which
is odd under both parities. The LMSP could be the LSP, the LMP,
both or neither one.
If both parities are preserved, the LSP and LMP are stable.
In addition, there could also be a third stable 
particle whose decay 
into the LSP and LMP is kinematically forbidden.
(For example, this could be the LMSP if it is not the LMP
or the LSP and it is not heavy enough to decay into them.) 
A stable charged (electromagnetic or 
color) particle is excluded up to $\sim {\cal{O}}(20\,
{\rm TeV})$ cosmologically from 
failure of terrestrial searches for anomalously-heavy  isotopes of various 
elements \cite{stable}. The only possible massive 
stable particles are therefore
the  mirror neutrinos; sneutrinos and mirror sneutrinos;
the Higgsinos, mirror Higgsinos and mirror Higgs; 
$\phi_{\gamma}$ and $\psi_{\gamma}$; and $\phi_{Z}$ and $\psi_{Z}$ 
(where we rotated the electroweak group to its SM basis).
Note that even if the LSP is not stable due to 
the broken $R_{P}$ (RPV), there could still exist a stable neutral LMP,
which could be the candidate for dark matter. 
%
  
Particle decays in the MN2SSM can be classified as follows.
Define
\begin{itemize}
\item
$(+,\ +)$ to denote SM particles 
(quarks $q$, leptons $l$, Higgs bosons $H$, and gauge bosons $g,\ W,\ B$),
\item
$(-,\ +)$ to denote ordinary supersymmetric particles 
(squarks $\tilde{q}$, sleptons $\tilde{l}$, Higgsinos 
$\tilde{H}$, and gauginos $\tilde{g},\ \tilde{W},\ \tilde{B}$), 
\item 
$(+,\ -)$ to denote SM mirror particles (mirror squarks $q'$, mirror sleptons 
$l'$, mirror Higgs bosons $H'$, and mirror gauge bosons $\phi_V$), 
\item
$(-,\ -)$ to denote mirror sparticles 
(mirror squarks $\tilde{q}'$, mirror sleptons 
$\tilde{l}'$, mirror Higgsinos $\tilde{H}'$, and mirror gauginos $\psi_V$), 
\end{itemize}
where the first sign in the parenthesis is the particle's $R_P$ charge, 
while the  second sign is its $M_P$ charge.  
The allowed two-body decays are then
\begin{eqnarray}
(+,\ +)
&
\longrightarrow&{(+,\ -)(+,\ -)},\ (-,\ +)(-,\ +),\nonumber \\
&&(-,\ -)(-,\ -),
\ (+,\ +)(+,\ +),\\
(-,\ +)
&\longrightarrow&{(-,\ -)(+,\ -)},\ (-,\ +)(+,\ +),\\
(+,\ -)
&\longrightarrow&{(-,\ -)(-,\ +)},\ (+,\ -)(+,\ +),\\
(-,\ -)
&\longrightarrow&{(-,\ +)(+,\ -)},\ (-,\ -)(+,\ +).
\end{eqnarray}
Three-body and many-body decays can be classified
by applying the two-body channels to off-shell processes.
(Quartic couplings need also to be considered it some cases.)
%

As a concrete example, 
consider a case with a stop $\tilde{t}$ as the lightest ordinary
sparticle, which is heavier than the LSP (taken to be 
the mirror sneutrino which is also the LMSP, as an example) 
and LMP (taken to be mirror neutrino).
A possible decay chain of stop is: 
$\tilde{t}\rightarrow\ {b}\tilde{W}
\rightarrow\ bl^{\prime}{\tilde\nu^{\prime}}
\rightarrow\ b{l}\phi_Z\tilde\nu^{\prime}
\rightarrow\ b{l}\nu\nu^{\prime}\tilde\nu^{\prime}$, 
if it is kinematically allowed.
(Intermediate steps could be on- or off-shell.) 
The final state contains in this case a $b$-jet, 
a charged lepton, and missing energy.
Alternatively, it could decay via a trilinear coupling:
$\tilde{t}\rightarrow\ \tilde{b}^{\prime}{{H}^{\prime}}^{+}$
leading to a similar final state 
$b{l}(3\nu)\nu^{\prime}\tilde{\nu}^{\prime}$.
The neutrinos and the neutral LMPs and LSPs all lead in this case 
to a missing energy signal, as in the usual $N=1$ MSSM case.  
%

Since the masses of the mirror (matter) fermions are related to the 
electroweak symmetry breaking scale and would be at most 
a couple hundred GeV,  they are most likely to be the first 
mirror particles to be produced at the 
colliders and are candidates for the LMP.
The mirror quarks particularly deserve attention.
They can be copiously pair produced at the LHC and the Tevatron 
via gluon fusion.  
If a mirror neutrino is the LMP and all the superparticles are heavier, 
each mirror quark can either decay through an on-shell 
electroweak mirror gauge boson:
$q_i'\rightarrow{q_i}\phi_{Z}, \ {q_i}\phi_{\gamma},\ {q_j}\phi_{W}$,  
where $\phi_{Z,\gamma}\rightarrow\nu\nu',\ \phi_{W}\rightarrow{l}\nu',$ 
if kinematically allowed; 
or it directly decays into $q_i\nu\nu', q_jl\nu'$ through an off-shell
process. A typical event has to two energetic jets 
(two charged leptons, in some cases) and a missing energy.
If the superparticles are not too heavy, mirror quarks can alternatively
decay through supersymmetric channels.  
In addition to the jets and leptons,  
the final state could have in this case at least two LSPs.
(The event reconstruction is  more difficult in this case.)
%

If one of the parities is violated, there is only one stable particle, and 
if both are violated, all the particles will eventually decay into 
SM jets and leptons. The LSP and LMP lifetimes depend in this case
on the extent of the respective parity violation.
If the LMP decays outside the detector,  
it appears in the detector as a stable particle.
Consider a mirror fermion as the (meta-stable) LMP.
A neutrino (or any other neutral)
LMP leads in this case as well to a  missing energy signal.
Mirror charged lepton LMPs leave a track in the central 
tracking chamber and hit the muon chamber, with less activity in the 
calorimeters.  A muon and a mirror charged lepton 
can be distinguished  by the ionization rate $dE/dx$ since the mirror 
particle is much heavier.  If a mirror quark
is the LMP and stable inside 
the detector, it will form a quarkonium or combine 
with the ordinary quarks to form a mirror hadron.  
Such states will lead to hadron 
showers in the hadron calorimeters, and can be distinguish
from a regular hadron shower by the wider shower opening angle \cite{hadron}.
If a mirror gauge boson is  the LMP  (and stable in the detector) 
the signals will be similar to those mentioned above, 
depending on whether it is neutral, charged and/or carrying color. 
If the LMP decays inside the detector, the heavy mirror particle can decay
into jets, leptons, or lighter supersymmetric particles.  A missing energy 
signal is still possible if the usual $R$-parity is exact.
Otherwise, the signal mimics those of 
the SM heavy fourth family quarks and leptons and of the RPV MSSM.  
%

We conclude that once the  center of mass energy of the future colliders 
is sufficient  to produce the mirror fermions, 
they can hardly escape detection.  
Since future colliders will effectively provide 
top factories, sufficient energies
will be reached, providing an ideal environment
for searching for the $N=2$ mirror quarks (which cannot be much heavier
than the top quark).
%

Higgs production is also affected by the MN2SSM spectrum.
The existence of extra heavy mirror  quarks can greatly enhance 
the single-Higgs production rate in hadron colliders through 
gluon and quark fusions, since the effective Yukawa coupling is of the 
order of unity. Neutral Higgs bosons can be produced radiatively  via 
$gg\rightarrow{H}^0$  through heavy quark loop \cite{higgs1}.  
Higgs-strahlung associate production rates
for both neutral \cite{higgs2} and charged \cite{higgs3} Higgs bosons 
through $2\rightarrow{3}$ processes $gg,\ qq\rightarrow{q^{\prime}
q^{\prime}H}$ (and in the case of MPV also 
$2\rightarrow{2}$ processes 
$qg\rightarrow{q^{\prime}}H $) can also increase
greatly.  For example, Ref.~\cite{tevatron} argued that  
one generation of mirror heavy quarks can increase the Higgs production cross 
section for $gg\rightarrow{H}^0$ by a factor of six to nine.
In Ref.~\cite{higgs2} the contributions of large
Yukawa couplings to associate Higgs production
was shown for the case of large $\tan\beta$ and of RPV.
In addition, decay channels of (a sufficiently heavy) Higgs to 
mirror fermions are also expected to be important.
Lastly, if there are radiative Yukawa couplings, they also affect
Higgs phenomenology and create misalignment between on-shell and mass
($m_{f}/\langle H \rangle$) Yukawa couplings of the Higgs \cite{BFPT}.
%

The sparticle phenomenology is also richer than in the MSSM.
For example, consider
the (eight Majorana) neutralinos and (four Dirac) charginos. 
If the mirror parity is unbroken, there are two diagonal blocks in the
neutralino/chargino mass matrices, each of which 
is an analogue of the usual $N=1$ case.  
The $M_{P}$-even neutralino block could provide
the LSP while that odd block could provide the LMSP
(that may or may not be the LMP and/or LSP). Each sector could decay, however,
to particles in the other sector via $M_{P}$ conserving couplings
such as $\widetilde{H}^{\prime}_{i}\phi_{Z}\widetilde{H}_{i}$.
If mirror parity is violated then there are off-diagonal
mixing terms between the ordinary and the mirror Higgsinos and gauginos,  
which lead to complicated mixing patterns.
Similarly, the mixing patterns of  squarks
and sleptons are also complicated in the presence of MPV terms, 
while new production and decay channels (and complicated cascades)
are open whether $M_{P}$ is conserved or violated.
%
%


%
\section{Conclusions}
\label{sec:conclusions}
%

In this paper we have formulated a low-energy $N=2$ supersymmetric framework
in which $N=2$ supersymmetry is preserved down to TeV energies.
The minimal $N=2$ realization of the SM, the MN2SSM, was presented
and its properties studied. While from the low-energy point of view
the models do not have a clear added benefit in comparison to
the $N=1$ MSSM to justify the
extended spectrum, it is ultra-violet constructions that suggest
the possibility of $N=2 \rightarrow N=0$ supersymmetry breaking \cite{breaking}. 
Therefore, it is important to examine the viability of such a scenario, 
address the issues it raises
(particularly its non-chiral nature), and investigate its signatures.
The next generation of hadron collider,
in which top pairs will be produced in abundance, is ideal
to test the $N=2$ framework via its mirror quark sector,
adding urgency to such an investigation.
To conclude, we review the main issues studied in this paper,
comment on some other issues such as unification and cosmology,
compare our framework to previous proposals, and propose further avenues for
investigation of the $N=2$ framework.
\subsection{The Framework}
%

$N=2$ supersymmetry was assumed in this work to break to $N=0$ at low-energies.
We chose, however, to formulate it as an $N=1$ theory, constrained by a set
of global $R$ symmetries which preserve the $N=2$ structure. 
The spectrum is that given by embedding the SM in $N=2$ 
hyper and vector multiplets, but it was written in terms of
their $N=1$ superfield components.
The minimal embedding corresponds to the MN2SSM:
Each MSSM superfield is accompanied
by a {\it mirror} superfield in the conjugate gauge representation.
The $N=2$ symmetries, in turn, constrain
the superpotential describing these superfield component interactions. 
In particular, an exchange symmetry between a particle
and its mirror and an Abelian $R$ symmetry imply that
the superpotential does not contain chiral Yukawa terms
and mass terms, respectively.
(Note that theory considered is a global $N=2$ described
in the $N=1$ language, i.e., gravity was not introduced.)
%

The $N=1$ language allows one to use the spurion formalism
and supersymmetry breaking is parameterized by two independent parameters,
the spurion auxiliary $F$-VEV and a mass parameter $M$ which suppresses
explicit breaking of the global symmetries in the Kahler potential,
with $F \simeq M^{2}$.
If supersymmetry is to play a role in resolution of the SM
hierarchy problem then $M \simeq {\cal{O}}({\rm TeV})$. Hence, supersymmetry
is broken at low energy and one has to consider all operators
to order $F^{2}/M^{4}$ in the Kahler potential. The effective theory
below the supersymmetry breaking scale
contains various quartic and Yukawa terms
and the $N=2$ and $N=1$ relations between couplings are broken.
Even though such breaking is hard, it does not destabilize the theory
but only affects the calculability of dimensionful parameters.
%

Below the supersymmetry breaking scale, some of the $N=2$ global 
symmetries may be preserved and could distinguish the SM matter from its 
mirror. In addition, parity symmetries, which do not commute with 
supersymmetry, may be admitted by the supersymmetry breaking mechanism. 
$R$-parity and mirror parity were used to define sparticles (as in the MSSM) 
and mirror particles, respectively. If preserved, each parity corresponds
to a stable neutral particle, the LSP and LMP, respectively.
\subsection{Fermion Mass Generation}
%

The explicit supersymmetry breaking terms in the Kahler potential
must also break the vectorial symmetries imposed by the 
$N=2$ global $R$-symmetries so that chiral Yukawa couplings
can be generated. This can be done by $N=1$ preserving, softly breaking,
or hard breaking tree-level operators $\sim (F/M^{2})^{n}$, $n=1,\,2.$
It can also be achieved by first breaking the symmetries by trilinear
terms in the scalar potential and generating the Yukawa terms 
at one loop. The latter leads to ${\cal{O}}$(GeV) quark masses
while the former, in principle, could lead to dangerous
mixing between the SM fermions and their mirrors.
However, by invoking the global ($R$) symmetries, one can distinguish
between operators which generate the SM fermion masses from
those generating the mirror spectrum so that the former
is proportional to suppressed Yukawa couplings or given by loop
corrections while the latter is given by ${\cal{O}}(1)$ Yukawa couplings.
In particular, a large mass hierarchy between the SM fermions 
and the mirror fermions can be achieved and the mixing between the sectors
can be suppressed or (if mirror parity is exact) even forbidden. 
The SM fermion spectrum is controlled, as usual, by small
Yukawa couplings while the mirror fermion spectrum
is constrained from above by the electroweak scale $m_{f^{\prime}}
\sim \langle H \rangle$. 
%

The heavy SM third family (especially the top quark) 
is an exception as its mass range is closer to that of the
mirror fermions than to that of the lighter SM generations.
Though such an hierarchy can be imposed by hand, here we considered
the possibility that a mirror symmetry also plays the role
of a SM flavor symmetry, allowing for
mirror parity violating mixing in the third family.
The  heavy top quark is then explained by the heaviness of the
third generation mirror top. A toy model with 
an Abelian flavor symmetry was given along these lines.
%

A very small violation of mirror parity may also play 
a role in the smallness of  neutrino
masses.  We suggested that the small neutrino masses can arise 
from a variation on the see-saw mechanism where the SM neutrino mass  
which is controlled by
the small mixing parameter between the ordinary and  mirror neutrinos.
The mirror neutrinos must be heavier 
than $m_Z/2$, as required by the invisible $Z$-width,
but could explain the smallness of the oblique parameter $S$
if they are not much heavier than that. This in fact occurs
in the same framework as the second eigenvalue is given by
$\sim \langle H \rangle^{2}/ M$. 
\subsection{Signals and Constraints}
%

The contributions to the oblique $S$-parameter from 
the extra three chiral generations provides a strong constraint
on the MN2SSM. In particular, it implies
that the mirror fermions cannot be degenerate in mass
and cannot (all) be too heavy, for example,
the mirror neutrino may be relatively light.
In addition, it also suggests light neutralinos and mirror
neutralinos (with significant custodial symmetry breaking mixing).
A dedicated fit to electroweak data in the MN2SSM framework
is necessary, but is beyond the scope of this work.
%

On the other hand, due to the low-energy supersymmetry breaking,
the Higgs mass is not constrained from above as in the MSSM
since its mass could be proportional to an arbitrary quartic couplings.
The Higgs sector could contain two  or four doublets,
any number of which could participate in supersymmetry breaking.
Choosing the MSSM limit of two ordinary Higgs doublets acquiring VEVs,
we find a Higgs potential which is more similar to that
of a generic two-Higgs doublet model than to the MSSM.
%

The spectrum of new particles is very rich, and contain
three new sectors (sparticles, mirror matter, mirror sparticles)
which do not mix with each other,
and two (in some cases even three) stable particles.
If either $R$ or mirror parity is broken, 
mixing is introduced and the number of stable neutral particles
is reduced respectively. The most obvious candidates for discovery
are the relatively light ${\cal{O}}(100-300)$ GeV mirror quarks,
while any stable particles are likely to correspond to missing energy.
%

The extra generations of mirror quarks and leptons at the 
electroweak scale provide the smoking gun for testing 
the $N=2$ framework at the LHC and the Tevatron.
Typical events consist of jets $+$ leptons $+$ missing 
(transverse) energy.  The missing energy is generally greater than 
in the $N=1$ supersymmetry ``events'', as the final states could 
includes two or more neutral heavy stable particles. 
(If mirror parity is slightly violated, 
there still is a stable particle as 
in the usual $N=1$ case, as long as $R$-parity is preserved, and vice versa.) 
Detail studies of the collider phenomenology
of low energy $N=2$ theories is also called upon, but clearly, 
it would be difficult for $N=2$ to escape discovery
if realized at TeV energies.
%

We note that 
aside from direct searches for mirror particles, extra 
generations of heavy quarks could greatly increase the Higgs 
production at hadron colliders
via gluon fusion, both single production
and  production in association with quarks are, in principle, enhanced.  
Therefore, an enhanced Higgs production cross section in hadron colliders
could be a hint for the existence of extra mirror matters.  
In addition, if the SM Yukawa couplings are radiative,
it carries strong implications to Higgs physics, such as misalignment 
of mass and on-shell Yukawa couplings, as well as to low-energy observables
such as an enhancement by an inverse loop factor
of the anomalous muon magnetic moment.
%

The phenomenological implications of $N=2$ supersymmetry
are rich. Here we focused on those which could be studied
in a relatively model-independent fashion.
However, the details of the model, for example, the
extent of mirror parity violation (if any) and the flavor
theory, can determine many aspects of the model phenomenology 
such as the stable particles, the cascade chains,
and indirect effects in low-energy and electroweak SM observables.
\subsection{Other Issues of Interest}
%

The high precision in which the SM gauge couplings are currently measured
strengthens the successful gauge coupling unification picture
in $N=1$ theories. (See, for example,  Ref.~\cite{unification}.)
Unfortunately, MSSM-like Planckian unification of the gauge couplings 
seems inconsistent with the framework of low-energy $N=2$.  
Since there are no higher-loop corrections in $N=2$ it is sufficient
to examine unification at one-loop order.
The one-loop beta function
coefficients of the SM gauge couplings in the MN2SSM 
are large and positive:
\begin{equation}
b_1^{N=2}=\frac{66}{5}, \ \ 
b_2^{N=2}=10, \ \ 
b_3^{N=2}=6.
\end{equation} 
Taking $M$ as before to be the $N=2$ breaking scale, above which one has 
the $N=2$ MN2SSM spectrum while between $m_{Z}$ and  $M$ resides either 
the $N=1$ MSSM or $N=0$ SM spectrum, we find
that $M$ has to be of the order of $10^{14}$ GeV 
(for $N=1$ below $M$ \cite{unifi}) or $10^{11}$ GeV (for $N=2$ 
breaking directly to $N=0$) for MSSM-like unification to hold.
This is inconsistent with the assumption that the $N=2$ 
breaking scale is near the electroweak scale.  (The situation
does not improve if there is only one Higgs doublet hypermultiplet.)
Alternatively, the MN2SSM  implies unification
$2-4$ orders of magnitude above its scale $M$, i.e.,
at intermediate energies.
Indeed, given the non-asymptotically
free behavior of the gauge couplings 
it is hard to imagine that there is a true desert 
between a low-energy $N=2$ breaking scale
and the sub-Planckian MSSM unification scale of $10^{16}$ GeV.
New physics may 
manifest itself as an extended gauge group, new thresholds,
or even an extended $N=4$ supersymmetry (which is finite),
and could play a role in resolving the unification issue.  
Note that the embedding in $N=4$ must involve
also extending the gauge group as all of the SM representations
must be embedded in that case in the adjoint representation (of an extended
gauge group), which is an interesting possibility.
%

Another issue of interest that we did not pursue here 
is the cosmological and astrophysical implications of such scenarios.
It is interesting to revisit issues such
as electroweak scale inflation \cite{inflation}
and electroweak scale baryogenesis \cite{bg} which can be sensitive
to the new rich electroweak structure.
In particular, we expect baryogenesis constraints
to be affected by the presence of large Yukawa couplings and the 
large number of Majorana fermions and of singlet fields \cite{bgsinglet}. 
A detailed study of neutrino mass and mixing patterns is also of 
great interest. (Here we only addressed the question of the overall
scale of neutrino masses.)
%

We also note in passing that 
the cosmological constant is zero in the $N=2$ limit.
If the leading contribution from supersymmetry breaking
is then of the order of $M^{8}/\mplanck^{4}$ (in particular,
the $M^{4}$ terms is canceled), where gravitational
corrections assumed to be suppressed by inverse powers of $\mplanck$,
it leads to values of the cosmological constant 
consistent to the currently preferred value of 
$\sim({10}^{-3}\,{\rm eV})^{4}$ \cite{constant}.
The MN2SSM can provide a natural realization of the
general arguments for such a framework \cite{murayama}.
\subsection{Previous Works}
%

Previous attempts to construct $N=2$ models were
based on either quantum correction
or $N=2$ gauge-Yukawa couplings \cite{west}.
The former was described in detail in section~\ref{sec:loop}. 
Although the masses for the usual quarks and leptons 
(except the third generation) can be generated at the right order of
magnitude, the mirror quarks and leptons are typically too light.
%

Realizing that the radiatively generated mass is not sufficient for the mirror 
fermions, it was proposed \cite{west} that
the mirror fermion mass may be generated at tree
level via the only Yukawa term $[Y\Phi_VX]_F$ in the superpotential 
if the SM $SU(2)_{L}$ is extended to $SU(4)_{LR}$.
The gauge group is  $SU(3)_{c}\times SU(4)_{LR}\times U(1)_{Y}$ 
(3-4-1) and the ordinary and mirror matter representations are
\begin{equation}
X=\left(
\begin{array}{c}
X_L\\\bar{Y}_R
\end{array}
\right)\sim({\bf 1}\,\, {\rm or} \,\, {\bf 3},{\bf 4}, Q_Y),\ \ \ 
Y=\left(
\begin{array}{c}
Y_L\\\bar{X}_R
\end{array}
\right)\sim({\bf 1}\,\,{\rm or}\,\, {\bf\bar{3}},{\bf \bar{4}}, -Q_Y), 
\end{equation}
where $X_L$, $Y_L$ are the ordinary and mirror $SU(2)_L$ doublets,
while  $\bar{X}_R$, $\bar{Y}_R$ are $SU(2)_R$ doublets and their mirrors.
Here $Q_Y$, $-Q_Y$ is the $U(1)_Y$
hypercharge of the superfields $X_L$ and $Y_L$, respectively.  
Note that conventional ordinary and mirror particles mix in $X$ and $Y$.  
By appropriately arranging of the parameter in the scalar potential,
the $SU(4)$ mirror gauge boson
$\Phi_4$ acquires off-diagonal VEVs, which gives masses to the mirror quarks
and leptons, while the usual matter fields are kept massless,
\begin{equation}
\mu^{\prime}_{XY} = \sqrt{2}Y\left(\begin{tabular}{cccc}
0&0&$\nu$&0\\
0&0&0&$\nu$\\
0&0&0&0\\
0&0&0&0
\end{tabular}\right)X.  
\end{equation}
%

One shortcoming of this approach is that the mirror  fermion masses are
constrained by the EWSB scale,
\begin{equation}
{\mu^{\prime}_{U}}^2+{\mu^{\prime}_{D}}^2=2M_W^2,
\end{equation} 
which is not consistent with experiment (based on fourth family searches).
Another crucial drawback is that 
all of the SM fermions have only loop masses,
which is unacceptable in the case of the top (and the $\tau$-lepton).
A realization of the 3-4-1 scenario was derived more recently
from a spontaneously broken $N=2$ supergravity \cite{GPZ}.
%

Our approach relies instead on low-energy supersymmetry 
and on the tree-level operators it induces.
It is more accommodating for embedding the SM
than either the pure loop approach or the 3-4-1 approach
and it retains a sufficient predictive power.
\subsection{Outlook}
%

In this paper we addressed some of the more fundamental issues
such as the scale of supersymmetry breaking, fermion mass generation,
constraints and discovery prospects. In each of this area 
there is clearly room for further study and explorations,
as indicated in the discussions above. In particular, a consistent
analysis of all constraints on the one hand, and of collider signals
(including the Higgs sector) on the other.
%

While we focused on mirror symmetries and on distinguishing
matter from mirror matter,
we only briefly touched upon the issue of flavor symmetries.
In fact, flavor symmetries may be entangled
with the mirror symmetries, rendering the heavy SM fermions
``more similar to'' the mirror fermions. This is a new paradigm
for flavor symmetries which offers new avenues for
construction of theories of flavor which were not yet explored.
%

Here we subscribed to an effective approach,
allowing the most general Kahler potential, which is
constrained only by the symmetries.
Ultimately, one would like
to derive a suitable Kahler potential (and find the necessary terms)
from a spontaneously broken $N=2$ supergravity theory, and perhaps
from a more fundamental theory. (This was done in Ref.~\cite{GPZ}
for the 3-4-1 framework of Ref.~\cite{west}.) 
The scale of the  more fundamental theory need also to be explored
and it is suggestive that
some new physics exists only a few orders of magnitude
above the supersymmetry breaking scale.
The embedding of $N=2$ extensions of the SM in theories with
extra large spatial dimensions (where $N=2$ naturally appears)
and/or strongly interacting string theories 
is also worth exploring.
%

All in all, the viability and richness of $N=2$
extensions of the SM introduce many questions worth pursuing,
from mirror quark searches to Kahler potential construction
in the case of low-energy supersymmetry breaking.
These are integral parts of the MN2SSM construction, but
their implications extend beyond it:
Search algorithms in models with two stable particles
and limits on the Higgs mass in $N=1$ models with low-energy supersymmetry 
breaking are examples of issues that are both central to the MN2SSM
phenomenology and extend beyond the $N=2$ framework
and should be explored both within and independently of the MN2SSM.
%
%
%


\acknowledgements

We thank Jon Bagger, Francesca Borzumati, Paul Langacker, Erich Poppitz,  
and Lisa Randall for discussions and comments. We also thank Jens Erler
for private communications regarding the oblique parameters
in the MN2SSM. This Work is 
supported by the US Department of Energy under cooperative research
agreement No.~DF--FC02--94ER40818.


\end{document}